\documentclass[journal]{IEEEtran}
\usepackage{cite}

\ifCLASSINFOpdf
  \usepackage[pdftex]{graphicx}
\else
\fi
\usepackage{amsmath}
\usepackage{amssymb}
\usepackage{amsthm}
\usepackage{bm}
\usepackage{mathalpha}
\usepackage{algorithm}
\usepackage{algpseudocode}
\ifCLASSOPTIONcompsoc
 \usepackage[caption=false,font=normalsize,labelfont=sf,textfont=sf]{subfig}
\else
 \usepackage[caption=false,font=footnotesize]{subfig}
\fi
\usepackage{url}

\usepackage{physics}
\usepackage{booktabs}
\usepackage{threeparttable}
\usepackage{color}
\usepackage{multirow}
\usepackage{stmaryrd}
\usepackage{makecell}

\usepackage{hyperref}

\SetSymbolFont{stmry}{bold}{U}{stmry}{m}{n}

\newtheorem{remark}{Remark}

\newtheorem{proposition}{Proposition}

\newtheorem*{criterion*}{Criterion}

\newcommand{\figref}[1]{Fig. \ref{#1}}
\newcommand{\tabref}[1]{Table \ref{#1}}
\newcommand{\alref}[1]{Algorithm \ref{#1}}

\newcommand{\secref}[1]{Section \ref{#1}}

\newcommand{\propref}[1]{Proposition \ref{#1}}

\newcommand{\scaleofsimulation}{1}

\begin{document}
%
\title{Tensor-Structured Bayesian Channel Prediction for Upper Mid-Band XL-MIMO Systems}
%
%
%

\author{Hongwei~Hou,~\IEEEmembership{Graduate~Student~Member,~IEEE},
  ~Yafei~Wang,~\IEEEmembership{Graduate~Student~Member,~IEEE}, \\
  ~Xinping Yi,~\IEEEmembership{Member,~IEEE},
  Wenjin~Wang,~\IEEEmembership{Member,~IEEE},\\
  ~Dirk~T.~M.~Slock,~\IEEEmembership{Life~Fellow,~IEEE},
  Shi~Jin,~\IEEEmembership{Fellow,~IEEE}
\thanks{}
\thanks{}
\thanks{}
\thanks{Hongwei Hou, Yafei Wang, and Wenjin Wang are with the National Mobile Communications Research Laboratory, Southeast University, Nanjing 210096, China, and also with Purple Mountain Laboratories, Nanjing 211100, China (e-mail: hongweihou@seu.edu.cn; wangyf@seu.edu.cn; wangwj@seu.edu.cn).}
\thanks{Xinping Yi and Shi Jin are with the National Mobile Communications Research Laboratory, Southeast University, Nanjing 210096, China (e-mail: xyi@seu.edu.cn; jinshi@seu.edu.cn).}
\thanks{Dirk T. M. Slock is with the Department of Communication Systems, EURECOM, 06410 Biot, France (e-mail: Dirk.Slock@eurecom.fr).}}

%
%

\markboth{ }%
{ }
%



\maketitle

\begin{abstract}%
    The upper mid-band balances coverage and capacity for the future cellular systems and also embraces extremely large-scale multiple-input multiple-output (XL-MIMO) systems, offering enhanced spectral and energy efficiency.
    However, these benefits are significantly degraded under mobility due to channel aging, and further exacerbated by the unique near-field (NF) and spatial non-stationarity (SnS) propagation in such systems.
    To address this challenge, we propose a novel channel prediction approach that incorporates dedicated channel modeling, probabilistic representations, and Bayesian inference algorithms for this emerging scenario.
    Specifically, we develop tensor-structured channel models in both the spatial-frequency-temporal (SFT) and beam-delay-Doppler (BDD) domains, which capture the NF and SnS propagation effects and leverage temporal correlations among multiple pilot symbols for channel prediction.
    In this model, the factor matrices of multi-linear transformations are parameterized by BDD domain grids and SnS factors, where beam domain grids are jointly determined by angles and slopes under spatial-chirp based NF representations.
    To enable tractable inference, we replace these environment-dependent BDD domain grids with uniformly sampled ones, and introduce perturbation parameters in each domain to mitigate grid mismatch.
    We further propose a hybrid beam domain strategy that integrates angle-only sampling with slope hyperparameterization to avoid the computational burden of explicit slope sampling.
    Based on the probabilistic models, we develop tensor-structured bi-layer inference (TS-BLI) algorithm under the expectation-maximization (EM) framework, which reduces the computational complexity through tensor operations.
    In the E-step, we develop the bi-layer factor graph representation to isolate the bilinear mixing in the spatial domain induced by SnS propagation, thus facilitating bi-layer iterations using approximate inference techniques.
    In the M-step, we leverage an alternating strategy for hyperparameter learning, with closed-form rules derived by the quadratic approximation of objective functions.
    Numerical simulations based on the near-practical channel simulator demonstrate the superior channel prediction performance of the proposed algorithm.
\end{abstract}%


%
\IEEEpeerreviewmaketitle


\section{Introduction}
\IEEEPARstart{A}{s} next-generation cellular communication systems strive to support data-intensive applications, the demand for spectrum that enables both high capacity and broad coverage has become increasingly critical \cite{giordani2020toward, na2024operator, wang2024towards}. 
In response, growing attention has turned to the newly defined frequency range (FR), with which FR3 (upper mid-band) has emerged as the most promising candidate that offers a favorable trade-off between the spectral congestion of FR1 (sub-$6$ GHz) and the severe propagation limitations of FR2 (millimeter-wave) \cite{3gpp38820, kang2024cellular}.

The relatively short wavelengths of FR3 make it feasible to deploy extremely large aperture arrays (ELAA) at the base station (BS) \cite{tian2025mid}, which  have been recognized as a key enabling technology to sustainably boost both spectral and energy efficiency.
To unleash this potential, the acquisition of channel state information (CSI) is essential, yet increasingly challenging under mobility due to channel aging.
This challenge is further exacerbated in FR3 systems, where the higher carrier frequencies induce more significant channel aging than in FR1, and the richer multipath propagation compared to FR2 renders straightforward Doppler compensation ineffective \cite{miao2023sub}.
Therefore, channel prediction has emerged as a promising solution by exploiting temporal correlations in historical CSI to predict future CSI.
However, the large physical aperture of ELAA brings scatterers and mobile terminals (MTs) within the Rayleigh distance, and also results in non-uniform visibility across the array \cite{yuan2023spatial}.
These effects give rise to unique channel characteristics in FR3, incorporating both near-field (NF) and spatial non-stationarity (SnS) propagations, which significantly reshape the beam domain channel representations and further complicate channel prediction. 

\subsection{Previous Works}
Before delving into the challenges specific to FR3 systems, it is instructive to review representative channel prediction techniques developed for massive multiple-input multiple-output (MIMO) systems. 
Among the earliest contributions, \cite{guillaud2004specular, adeogun2015extrapolation} explore channel prediction using spectral estimation techniques and parametric channel models, where \cite{adeogun2015extrapolation} also derives performance bounds that quantify the theoretical limits of channel prediction. 
In the context of massive MIMO with orthogonal frequency-division multiplexing (OFDM), \cite{yin2020addressing, qin2022partial, li2022multi} extract Doppler frequencies from dominant angle-delay domain channel taps, while \cite{wu2021channel} adopts autoregressive (AR) modeling as an alternative approach, with both achieving significant gains.
For more practical channel dynamics, \cite{wang2023channel, wan2023robust, wan2024two} incorporate time-varying Doppler frequencies through polynomial expansions and AR processes, as exemplified by \cite{wang2023channel} through validation on measured channel data.
To alleviate the effect of channel estimation error, \cite{zhu2024joint, shi2022channel, hou2024tensor} unify the channel estimation and prediction based on beam-delay-Doppler (BDD) domain representations, and achieve the minimum mean square error (MMSE)-based channel prediction.

Despite extensive research on channel prediction in massive MIMO systems, this topic remains largely unexplored in the context of extremely large-scale MIMO (XL-MIMO) equipped with ELAA.
Most existing works focus on channel estimation or tracking, which aim to reconstruct the currently observed channels rather than to predict future, unobserved ones.
Specifically, \cite{cui2022channel, lu2023near} propose polar domain representations based on the angle-distance sampling to exploit the inherent sparsity under NF propagation, where \cite{cui2022channel} introduces grid refinement to mitigate quantization errors. 
Besides the NF propagation, hybrid-field models are adopted in \cite{9940281, 9598863, hou2024beam}, jointly incorporating angle and angle-distance domain representations.

Besides NF propagation, recent studies have focused on capturing SnS, which has emerged as another fundamental channel characteristic in XL-MIMO systems.
Assuming local spatial stationarity within small subarrays, \cite{han2020channel} and \cite{chen2024non} partition the ELAA into uniform sub-arrays, while \cite{yang2025adaptive} proposes an adaptive segmentation strategy to enhance modeling accuracy.
In contrast, \cite{zhu2021bayesian} directly models the antenna domain sparsity induced by SnS, focusing exclusively on the structural support of active elements. 
Building upon this, \cite{tang2024joint, xu2024joint, xu2025exploiting, tang2024spatially} explicitly incorporate inter-antenna correlations caused by multipath propagation using beam domain representations.
For modeling simplification, \cite{tang2024joint} and \cite{tang2024spatially} assume shared visibility regions across all paths and directional beam-based sub-channel decomposition, respectively, while \cite{xu2024joint} and \cite{xu2025exploiting} design multi-module architectures that alternate between beam domain channel and SnS factor inference.
Most recently, \cite{li2024wavefront} presents the first attempt to address channel prediction in XL-MIMO systems, where the spherical wavefronts are approximated as near-planar ones based on wavefront transformations (WT), and Doppler frequencies are extracted by the WT matrix pencil (WTMP) approach to facilitate channel prediction.

\subsection{Motivations and Contributions}
In upper mid-band XL-MIMO systems, the higher carrier frequency than FR1 and the richer multipath than FR2 bring the increased Doppler frequencies and spread, which emphasizes the necessity of channel prediction in combating more severe channel aging.
However, the deployment of ELAA introduces the coexistence of NF and SnS propagation, rendering the exploitation of inter-antenna correlation challenging and further degrading the channel prediction performance.
Despite these challenges, existing efforts remain limited—either ignoring SnS propagation or restricting attention to the channel estimation task—thereby leaving a critical gap in channel prediction for upper mid-band XL-MIMO systems.

To bridge this gap, we investigate tensor-structured Bayesian channel prediction for upper mid-band XL-MIMO systems in this paper, pioneering the integration of NF and SnS propagation effects into channel prediction. The main contributions of this work are summarized as follows:
\begin{itemize}
    \item Within a fixed-length observation window, we develop the tensor-structured spatial-frequency-temporal (SFT) domain and BDD domain channel models with parameterized factor matrices, incorporating both NF and SnS propagation effects. 
    Specifically, the spatial domain factor matrices are parameterized by instantaneous angles and variation rates (i.e., slopes) under the spatial-chirp representation of NF propagation, while antenna-dependent visibility factors capture the SnS propagation.
    This modeling framework enables arbitrary-length channel prediction through the transformation from the Doppler domain to the temporal domain, while the beam and delay domains further enhance the channel prediction by capturing the inter-antenna and inter-subcarrier correlations.
    \item To enable tractable inference, we propose dedicated probabilistic models for channel prediction, which capture multi-linear transformations, BDD domain channel and SnS factor sparsity. 
    Specifically, we sample the BDD domain uniformly and introduce perturbation parameters in each domain to alleviate grid mismatch, thereby replacing dynamic grids in multi-linear transformations by fixed ones.
    We further adopt a hybrid beam domain sampling strategy to bypass the increased computational complexity due to extra slope sampling under NF propagation, which integrates angle-only sampling and slope hyperparameterization to strikes the balance between model complexity and expressive capacity.
    \item Following these probabilistic representations, we propose the tensor-structured bi-layer inference (TS-BLI) algorithm within the expectation-maximization (EM) framework for channel prediction.
    Specifically, we design a bi-layer factor graph in the E-step of the EM framework to isolate the bilinear mixing in the spatial domain induced by SnS propagation, thereby deriving tractable iterations via approximate inference techniques.
    In the M-step, the alternating optimization strategy is employed to shrink the search space of hyperparameters, while approximating the objective functions yields near-optimal closed-form update rules.
    The proposed tensor-structured inference not only inherently supports the decoupling of linear and bilinear mixing, but also reduces computational complexity through efficient tensor operations.
\end{itemize}

\subsection{Organization and Notations}
\subsubsection{Organization}
The SFT domain channel model and its Tucker-based BDD domain representations are developed in \secref{sec:system_model}.
Based on the probabilistic models, we formulate the MMSE-based channel prediction problem and propose the corresponding TS-MLI algorithm in \secref{sec:problem_formulation} and \secref{sec:online_TSDCP}, respectively.
\secref{sec:simulation_result} presents the numerical simulations, and \secref{sec:conclusion} concludes the paper.

\subsubsection{Notations}
The imaginary unit is represented by $j = \sqrt{-1}$. 
$x$, $\mathbf{x}$, and $\mathbf{X}$ denote scalars, column vectors, and matrices, respectively. 
The transpose, conjugate, and conjugate-transpose operations are represented by the superscripts $(\cdot)^{T}$, $(\cdot)^{\ast}$, and $(\cdot)^{H}$, respectively.
The symbols $\mathbb{C}$ denote the complex number fields.
$[\cdot]_{i_{1}, {\dots}, i_{D}}$ is the $(i_{1}, {\dots}, i_{D})$-th element of $D$-order tensor.
$\mathsf{D}\left\{\cdot\right\}$ and $\mathsf{H}\left\{\cdot\right\}$ denote the Kullback-Leibler (KL) divergence and differential entropy, respectively.
The outer product, element-wise multiplication, and division are denoted by $\circ$, $\odot$, and $\oslash$ respectively.
$\mathsf{E}\left\{\cdot\right\}$ and $\mathsf{V}\left\{\cdot\right\}$ denote the expectation and variance operators, respectively.
$\mathsf{diag}\{\cdot\}$ and $\mathsf{Re}\{\cdot\}$ denote the diagonal and real part operators, respectively.

\subsubsection{Tensor Notations}
The tensor operations and definitions in this paper align with the counterparts in \cite{kolda2009tensor}. For a $D$-order tensor $\boldsymbol{\mathcal{X}}{\;\in\;}\mathbb{C}^{N_{1}{\times}N_{2}{\times}{\dots}{\times}N_{D}}$, 
the mode-$d$ matrixization $\mathbf{X}_{d}{\;\in\;}\mathbb{C}^{N_{d}{\times}N_{1}{\dots}N_{d-1}N_{d+1}{\dots}N_{D}}$ arranges the mode-$d$ fibers of this tensor into its column vectors, obtained by fixing the index along the $d$-th mode and varying the others.
Given the tensors $\boldsymbol{\mathcal{X}}, \boldsymbol{\mathcal{Y}}$ with the same size, the inner product is defined as
\begin{equation}
  \langle \boldsymbol{\mathcal{X}}, \boldsymbol{\mathcal{Y}} \rangle = \sum_{n_{1}}\sum_{n_{2}}{\dots}\sum_{n_{D}} [\boldsymbol{\mathcal{X}}]_{n_{1}, n_{2}, {\dots}, n_{D}}[\boldsymbol{\mathcal{Y}}]_{n_{1}, n_{2}, {\dots}, n_{D}}^{\ast},
\end{equation}
which is also the high-order extension of matrix inner product, and we define $\|\boldsymbol{\mathcal{X}}\|_{F} = \sqrt{\langle \boldsymbol{\mathcal{X}}, \boldsymbol{\mathcal{X}} \rangle}$ as the high-order extension of Frobenius norm.
We also define $\ell_{1}$ norm of $\boldsymbol{\mathcal{X}}$, given by $\|\boldsymbol{\mathcal{X}}\|_{1} = \sum_{n_{1}}\sum_{n_{2}}{\dots}\sum_{n_{D}} |[\boldsymbol{\mathcal{X}}]_{n_{1}, n_{2}, {\dots}, n_{D}}|$.
The mode-$d$ tensor-matrix multiplication of tensor $\boldsymbol{\mathcal{X}}$ and matrix $\mathbf{U}_{d}{\;\in\;}\mathbb{C}^{K_{d}{\times}N_{d}}$ is denoted as $\boldsymbol{\mathcal{Y}} = \boldsymbol{\mathcal{X}}{\;\times_{d}\;}\mathbf{U}_{d}$, expressed as
\begin{equation}
  [\boldsymbol{\mathcal{Y}}]_{n_{1},{\dots}, n_{d-1}, k_{d}, n_{d+1},{\dots},n_{D}} = \sum_{n_{d}}[\mathbf{U}]_{k_{d},n_{d}}[\boldsymbol{\mathcal{X}}]_{n_{1},{\dots},n_{D}},
\end{equation}
and equivalent to $\mathbf{Y}_{d} = \mathbf{U}_{d}\mathbf{X}_{d}$.
In analogy with the Einstein product \cite{brazell2013solving}, we define the special case of tensor-tensor multiplications for $\boldsymbol{\mathcal{X}}{\;\in\;}\mathbb{C}^{M_{1}{\times}{\dots}{\times}M_{d-1}{\times}P{\times}M_{d+1}{\times}{\dots}{\times}M_{D}}$ and $\boldsymbol{\mathcal{Y}}{\;\in\;}\mathbb{C}^{M_{1}{\times}{\dots}{\times}M_{d-1}{\times}Q{\times}M_{d+1}{\times}{\dots}{\times}M_{D}}$ as $\mathbf{Z} = \boldsymbol{\mathcal{X}}{\;\times_{-d}\;}\boldsymbol{\mathcal{Y}}$. 
This operation applies to all modes except for the $d$-th one and is equivalent to the multiplication of their mode-$d$ matricizations, i.e., $\mathbf{Z} = \mathbf{X}_{d}\mathbf{Y}_{d}^{T}$.
$\mathsf{CN}( \boldsymbol{\mathcal{X}}; \boldsymbol{\mathcal{U}}, \boldsymbol{\mathcal{E}})$ and $\mathsf{BG}( \boldsymbol{\mathcal{X}}; \boldsymbol{\mathcal{M}}, \boldsymbol{\mathcal{U}}, \boldsymbol{\mathcal{E}})$ denote the joint probability density functions (PDFs) of $\boldsymbol{\mathcal{X}}$, whose entries are independently drawn from Gaussian distributions with element-wise mean $\boldsymbol{\mathcal{U}}$ and variance $\boldsymbol{\mathcal{E}}$, and Bernoulli-Gaussian distributions with element-wise sparsity $\boldsymbol{\mathcal{M}}$, mean $\boldsymbol{\mathcal{U}}$, and variance $\boldsymbol{\mathcal{E}}$, respectively.
$\boldsymbol{\mathcal{C}}(x)$ denotes the constant tensor with all elements being $x$.

\section{System Model}\label{sec:system_model}
\begin{figure}[!t]
  \centering
  \includegraphics[width = \linewidth]{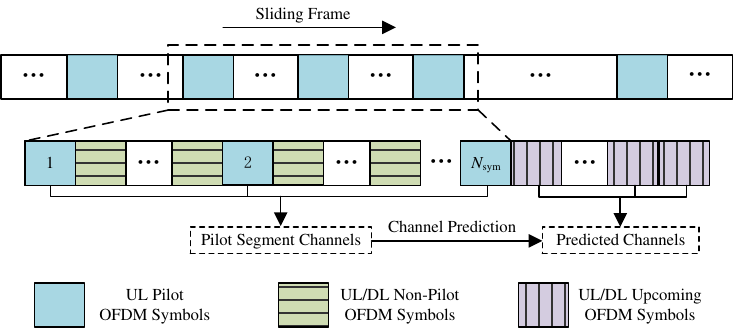}
  \caption{Frame structure of massive MIMO-OFDM systems in TDD mode.}
  \label{fig:FrameStructure}
\end{figure}

In this paper, we consider the upper mid-band XL-MIMO-OFDM system under the time-division duplexing (TDD) mode with one BS and multiple single-antenna MTs, where the BS is equipped with a uniform linear array (ULA) comprising $N_\text{an} {\;\gg\;} 1$ antennas. 
The OFDM symbol duration, cyclic prefix (CP) duration, and subcarrier spacing are denoted by ${\Delta}T_{\text{sym}}$, ${\Delta}T_{\text{cp}}$, and ${\Delta}f$ respectively, leading to the total OFDM symbol duration with CP given by ${\Delta}T = {\Delta}T_{\text{sym}} + {\Delta}T_{\text{cp}}$.
The MTs employ sounding reference signal as pilots for uplink channel sounding \cite{3gpp38211}, which is characterized by comb-type pilot patterns with uniform spacing in both temporal and frequency domains. Specifically, pilot symbols are inserted every $ N_\text{IS}$ symbols and $N_\text{TC}$ subcarriers in the temporal and frequency domains, respectively, corresponding to pilot spacings of ${\Delta}\bar{T} = N_\text{IS} {\Delta}T$ and ${\Delta}\bar{f} = N_\text{TC} {\Delta}f$.
With this configuration, we introduce the sliding frame structure shown in \figref{fig:FrameStructure}, where each frame contains multiple pilot OFDM symbols. 
In each frame, the BS collects the symbols in the pilot segment and predicts channels of all symbols from the last observed pilot symbol to the first upcoming pilot symbol, thus combating channel aging.

\subsection{Signal and Channel Models}
Based on ray-tracing models, the channel impulse response (CIR) at the $n_{\text{an}}$-th antenna consists of $L$ paths, expressed as\footnote{Since the comb-type pilot pattern eliminates pilot contamination, we focus on the typical MT and omit the MT index for brevity.}
\begin{equation}
    h_{n_{\text{an}}}(t, {\tau}) = \sum_{l=1}^{L}  {\beta}_{l} \tilde{s}_{l, n_\text{an}} \mathsf{exp}(j2{\pi}t\tilde{\nu}_{l}) \delta({\tau} - \tilde{\tau}_{l} -{\Delta}\tilde{\tau}_{l, n_{\text{an}}}),
\end{equation}
where ${\beta}_{l}$, $\tilde{\nu}_{l}$, and $\tilde{\tau}_{l}$ denote complex gain, Doppler frequency and reference propagation delay of the $l$-th path, respectively, $\tilde{s}_{l, n_\text{an}}$ and ${\Delta}\tilde{\tau}_{l, n_{\text{an}}}$ denote SnS factor and propagation delay difference of the $l$-th path and the $n_{\text{an}}$-th antenna\footnote{We consider only phase variations induced by spherical wavefronts. This approximation holds when the propagation distance exceeds ${\zeta}D/(2\sqrt{1-{\zeta}^2})$ \cite{sherman1962properties}, where $D$ is the array aperture and $\zeta$ is amplitude variation thresholds.}, respectively. 

\begin{figure}
    \centering
    \includegraphics[width = 0.9 \linewidth]{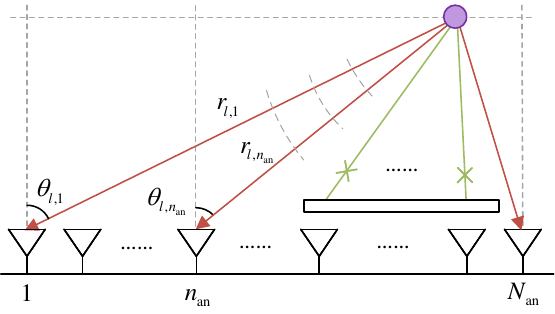}
    \caption{NF and SnS propagations in the upper mid-band XL-MIMO systems.}
    \label{fig:Spherical_Wave}
\end{figure}

To further characterize channels in upper mid-band XL-MIMO systems, we specify the definitions of $\tilde{s}_{l, n_\text{an}}$ and ${\Delta}\tilde{\tau}_{l, n_{\text{an}}}$ in the channel model. Specifically, the deployment of ELAA leads to partial array visibility due to limited-size objects, which may either obstruct the signal or act as partial scatterers, as shown by the last few antennas in \figref{fig:Spherical_Wave}. 
This gives results in antenna-dependent path visibility, thereby motivating the introduction of SnS factors. The channel measurements from both academia and industry have demonstrated that binary-valued SnS factors are sufficient to capture this essential features \cite{yuan2023spatial, 3gppr12406199}, thereby setting $\tilde{s}_{l, n_\text{an}} {\;\in\;} \{0, 1\}$.
Besides, ELAA introduces non-linear dependencies of the propagation delay on the antenna index due to NF propagation. Based on the geometric relations in \figref{fig:Spherical_Wave}, the propagation distance difference between the $n_\text{an}$-th antenna and the first antenna to the last-hop scatterer of the $l$-th path satisfies\footnote{We retain up to the second-order term and ignore the high-order ones, which are valid for propagation distances exceeding $D^{4/3}/(2{\lambda}^{1/3})$ \cite{selvan2017fraunhofer}.}
\begin{align}\label{eq:distance_approximation}
  \tilde{r}_{l, n_\text{an}} & \mathop{=}^{\left(a\right)} \sqrt{\left[\tilde{r}_{l, 1} \cos \tilde{\theta}_{l, 1}\right]^{2}\!+\!\left[\tilde{r}_{l, 1} \sin \tilde{\theta}_{l, 1}\!-\!(n_{\text{an}}\!-\!1)d\right]^{2}} \nonumber \\
  & \mathop{\;\approx\;}^{\left(b\right)} \tilde{r}_{l, 1}\!-\!(n_{\text{an}}\!-\!1)d\tilde{\phi}_{l, 1}\!+\!(n_{\text{an}}\!-\!1)^{2}d^{2}\frac{1\!-\!\tilde{\phi}_{l, 1}^{2}}{2\tilde{r}_{l, 1}},
\end{align}
where $\left(a\right)$ and $\left(b\right)$ follow from the constant geometric constraint $\tilde{r}_{l, n}\cos \tilde{\theta}_{l, n}$ and the second-order Taylor series of $\sqrt{1+x}$ at $x=0$, respectively, $\tilde{\phi}_{l, 1}{\;\triangleq\;}\sin \tilde{\theta}_{l, 1}$ and $\tilde{r}_{l, 1}$ denote the directional cosine and propagation distance of the $l$-th path, referred to as $\tilde{\phi}_{l}$ and $\tilde{r}_{l}$, respectively, and $d$ denotes the inter-antenna spacing. With such definition, we have 
\begin{equation}
    {\Delta}\tilde{\tau}_{l, n_\text{an}} = [-(n_{\text{an}}-1)d\tilde{\phi}_{l} + (n_{\text{an}}-1)^{2}d^{2}\tilde{\eta}_{l}]/c,
\end{equation}
where $\tilde{\eta}_{l} {\;\triangleq\;} (1-\tilde{\phi}_{l}^{2})/(2\tilde{r}_{l}) $ denotes slope parameters of the $l$-th path, characterizing the deviation of the spherical wavefront from the planar one under NF propagation. 
In this context, the term ``slope'' stems from the spatial-chirp characterization of the array response, which captures the variation of instantaneous direction cosine with respect to the antenna index.

Due to the sufficiently short duration of OFDM symbols, CIRs are assumed to be constant within one OFDM symbol. By Fourier transforming and stacking the CIRs of all antennas and pilot resource elements, the spatial-frequency-temporal (SFT) domain channel $\boldsymbol{\mathcal{H}}{\;\in\;} \mathbb{C}^{N_\text{an}{\times}N_\text{sc}{\times}N_\text{sym}}$ is given by
\begin{align}\label{eq:channel_model_cp}
    \boldsymbol{\mathcal{H}} = \sum_{l=1}^{L} {\beta}_{l} (\tilde{\mathbf{s}}_{l} {\;\odot\;} \mathbf{a}_{\text{SS}}(\tilde{\phi}_{l}, \tilde{\eta}_{l})) {\;\circ\;} \mathbf{b}(\tilde{\tau}_{l}) {\;\circ\;} \mathbf{c}(\tilde{\nu}_{l}),
\end{align}
where $\tilde{\mathbf{s}}_{l} = [\tilde{s}_{l, 1}, {\dots}, \tilde{s}_{l, N_\text{an}}]^{T}{\;\in\;}\mathbb{C}^{N_\text{an}{\times}1}$ denotes the SnS factor of the $l$-th path, $\mathbf{a}_{\text{SS}}({\phi}, {\eta}){\;\in\;}\mathbb{C}^{N_\text{an}{\times}1}$ denotes beam domain steering vector without SnS propagation effect, defined as $[\mathbf{a}_{\text{SS}}({\phi}, {\eta})]_{n_\text{an}}=\mathsf{exp}(j2{\pi}(n_{\text{an}}-1)d[\phi_{l} - (n_{\text{an}}-1)d\eta_{l}]/{\lambda})$ with ${\lambda} = c/f_\text{c}$, $f_\text{c}$, and $c$ being the wavelength, carrier frequency, and speed of light, respectively, $\mathbf{b}({\tau}){\;\in\;}\mathbb{C}^{N_\text{sc}{\times}1}$ and $\mathbf{c}({\nu}){\;\in\;}\mathbb{C}^{N_\text{sym}{\times}1}$ denote delay and Doppler domain steering vectors, defined as $[\mathbf{b}({\tau})]_{n_\text{sc}} = \mathsf{exp}(-j2{\pi}(n_\text{sc}-1){\Delta}\bar{f}{\tau})$ and $[\mathbf{c}({\nu})]_{n_\text{sym}} = \mathsf{exp}(j2{\pi}(n_\text{sym}-1){\Delta}\bar{T}{\nu})$, respectively, and we also define the beam domain steering vector with SnS propagation effect as $\mathbf{a}(\phi, \eta, \mathbf{s}) {\;\triangleq\;} \mathbf{a}_{\text{SS}}({\phi}, {\eta}) {\;\odot\;} \mathbf{s}$.

After the cyclic prefix removal and OFDM demodulation, the received signal at the pilot segment is expressed as
\begin{equation}\label{eq:observation}
    \boldsymbol{\mathcal{Y}} = \boldsymbol{\mathcal{X}} {\;\odot\;} \boldsymbol{\mathcal{H}} + \boldsymbol{\mathcal{Z}},
\end{equation}
where $\boldsymbol{\mathcal{X}}$ and $\boldsymbol{\mathcal{Z}}$ denote the pilot tensor and the additive white Gaussian noise at pilot segments, respectively. Since the pilot symbols are known at both BS and MTs, we assume that $\boldsymbol{\mathcal{X}}$ is an all-one tensor without loss of generality.

\begin{remark}
    Due to the inherent spatial consistency of wireless channels, we assume that the physical parameters of each path, including BDD domain grids and SnS factors, remain unchanged over the duration of interest.
\end{remark}

\subsection{Tucker-Based Representation}
Due to the limited scattering propagation environment, the SFT domain channel typically exhibits a low-rank structure, characterized by the sparsity of the core tensor in its Tucker-based representations.
To leverage this sparsity, we construct the factor matrices $\mathbf{A}({\bm{\phi}}, {\bm{\eta}}, {\mathbf{S}}){\;\in\;}\mathbb{C}^{N_\text{an}{\times}K_\text{be}}$, $\mathbf{B}({\bm{\tau}}){\;\in\;}\mathbb{C}^{N_\text{sc}{\times}K_\text{de}}$, and $\mathbf{C}({\bm{\nu}}){\;\in\;}\mathbb{C}^{N_\text{sym}{\times}K_\text{do}}$ of SFT domain channels based on the beam-delay-Doppler (BDD) domain grids and the corresponding SnS factors, defined as
\begin{subequations}
    \begin{gather}
         \mathbf{A}({\bm{\phi}}, {\bm{\eta}}, {\mathbf{S}}) = [ \mathbf{a}({\phi}_{1}, {\eta}_{1}, {\mathbf{s}}_{1}), {\dots}, \mathbf{a}({\phi}_{K_\text{be}}, {\eta}_{K_\text{be}}, {\mathbf{s}}_{K_\text{be}}) ],\\
         \mathbf{B}({\bm{\tau}}) = [ \mathbf{b}({\tau}_{1}), {\dots}, \mathbf{b}({\tau}_{K_\text{de}}) ],\\
         \mathbf{C}({\bm{\nu}}) = [ \mathbf{c}({\nu}_{1}), {\dots}, \mathbf{c}({\nu}_{K_\text{do}}) ],
    \end{gather}
\end{subequations}
where $K_\text{be}$, $K_\text{de}$, and $K_\text{do}$ denote the number of BDD domain grids, respectively, which are set to balance model expressiveness and complexity, ${\bm{\phi}} = [{\phi}_{1}, {\dots}, {\phi}_{K_\text{be}}]^{T}$, ${\bm{\eta}} = [{\eta}_{1}, {\dots}, {\eta}_{K_\text{be}}]^{T}$, ${\bm{\tau}} = [{\tau}_{1}, {\dots}, {\tau}_{K_\text{de}}]^{T}$, and ${\bm{\nu}} = [{\nu}_{1}, {\dots}, {\nu}_{K_\text{do}}]^{T}$ denote the environment-dependent dynamic grids in the BDD domain, ${\mathbf{S}} = [{\mathbf{s}}_{1}, {\dots}, {\mathbf{s}}_{K_\text{be}}]{\;\in\;}\mathbb{C}^{N_\text{an}{\times}K_\text{be}}$, and ${\mathbf{s}}_{k_\text{be}}$ denotes the SnS factors corresponding to different beam domain grids.

With the aid of these factor matrices, the SFT domain channel is represented by the Tucker model, as given by
\begin{equation}\label{eq:multi-linear_channel}
    \boldsymbol{\mathcal{H}} = \boldsymbol{\mathcal{G}} {\;\times_{1}\;} \mathbf{A}({\bm{\phi}}, {\bm{\eta}}, {\mathbf{S}}) {\;\times_{2}\;} \mathbf{B}(\bm{\tau}) {\;\times_{3}\;} \mathbf{C}(\bm{\nu}),
\end{equation}
where $\boldsymbol{\mathcal{G}}{\;\in\;}\mathbb{C}^{K_\text{be}{\times}K_\text{de}{\times}K_\text{do}}$ denotes the BDD domain channel under these factor matrices.
The factor matrices $\mathbf{A}({\bm{\phi}}, {\bm{\eta}}, {\mathbf{S}})$, $\mathbf{B}(\bm{\tau})$, and $\mathbf{C}(\bm{\nu})$ represent the linear transformations from beam, delay, and Doppler domains to spatial, frequency, and temporal domains, respectively, which collectively constitute the multi-linear transformation from the BDD domain channel to the SFT domain channel.

Following this model, channel prediction is achieved by the acquisition of BDD domain channels, BDD domain grids, and corresponding SnS factors, followed by the multi-linear transformation defined in \eqref{eq:multi-linear_channel}.
With such physics-inspired model, the transformation from the Doppler domain to the temporal domain enables the arbitrary-length channel prediction, while the beam and delay domains capture the inter-antenna and inter-subcarrier correlations to further enhance the channel prediction performance.

\begin{remark}
    The prior work can be viewed as special cases of the proposed model under specific conditions. When we have $\mathbf{S} = \mathbf{1}_{N_\text{an}{\times}K_\text{be}}$, the representation degenerates to spatial stationary model in \cite{li2024wavefront} without delay variations. Moreover, by setting $\bm{\eta} = \mathbf{0}$, the representation further degenerates into the model in the conventional massive MIMO systems, which is consistent with the temporal stationary case in \cite{hou2024tensor}.
\end{remark}

\section{Problem Formulation}\label{sec:problem_formulation}

In the Tucker-based representations, prior knowledge of sparsity is essential for BDD domain channel acquisition, yet it is typically unknown or costly to obtain. 
This motivates revisiting it from the Bayesian perspective, which inherently enables automatic sparsity determination \cite{cheng2022towards}. 
Toward this end, we develop probabilistic models for the observation, multi-linear transformation, BDD domain channel, and SnS factor models, thus formulating the channel prediction problem.

\subsection{Probabilistic Model}\label{sec:problem_formulation_probabilistic_model}
\subsubsection{Observation and Multi-linear Transformation}
Based on the signal model in \eqref{eq:observation}, the observation model is given as
\begin{equation}
    \mathsf{P}(\boldsymbol{\mathcal{Y}} \!\mid\! \boldsymbol{\mathcal{H}}) {\;\propto\;} \mathsf{CN}(\boldsymbol{\mathcal{Y}}; \boldsymbol{\mathcal{H}}, \boldsymbol{\mathcal{C}}({\sigma}_{z}^{2})).
\end{equation}

To facilitate the inference of BDD domain channels, BDD domain grids, and SnS factors with given observations, it is essential to develop probabilistic models for multi-linear transformations defined in \eqref{eq:multi-linear_channel}. 
However, the environment-dependent dynamic grids in the multi-linear transformation complicates this model, which motivates the alternative of fixed grids to dynamic ones for tractable Bayesian inference. 
To guarantee sufficient coverage of the BDD domain, we build fixed grids via uniform sampling, yet challenges persist due to the NF propagation effect in the beam domain.
Specifically, the spatial-chirp characterization of the array response necessitates joint sampling over both angle and slope parameters, which significantly increases computational burden due to the enlarged spatial domain factor matrix. 
To mitigate this burden, we propose a simplified yet flexible strategy: only angles are sampled in the beam domain, with slope parameters treated as learnable hyperparameters. 
By doing so, we can significantly boost inference efficiency while preserving sufficient modeling capacity to capture NF propagation.

Despite the fixed grid enabling tractable Bayesian inference, such discretization in the BDD domain inherently limits the ability to capture continuous physical parameters. 
Instead of employing densely sampled grids directly, we incorporate perturbation parameters into the coarsely sampled ones to capture deviations from the ground-truth physical parameters.
Specifically, the perturbed parameters for angles, delays, and Doppler frequencies are expressed as
${\bm{\phi}} {\;\triangleq\;} \bar{\bm{\phi}} + \bm{\Delta}\bm{\phi} $, ${\bm{\tau}} {\;\triangleq\;}  \bar{\bm{\tau}} + \bm{\Delta}\bm{\tau} $, and ${\bm{\nu}} {\;\triangleq\;}  \bar{\bm{\nu}} + \bm{\Delta}\bm{\nu} $, respectively, where $\bar{\bm{\phi}}$, $\bar{\bm{\tau}}$, and $\bar{\bm{\nu}}$ denote coarsely sampled grids, while $\bm{\Delta}\bm{\phi}$, $\bm{\Delta}\bm{\tau}$, and $\bm{\Delta}\bm{\nu}$ denote the corresponding perturbation parameters that refine the discretizations.
With such sampling-then-refinement strategy, the PDF of multi-linear transformation is given as
\begin{align}
    \mathsf{P} ( \boldsymbol{\mathcal{H}} \mid \boldsymbol{\mathcal{G}}, {\mathbf{S}}; \;& \bm{\Delta}\bm{\phi}, \bm{\Delta}\bm{\tau}, \bm{\Delta}\bm{\nu}, {\bm{\eta}}) {\;\propto\;} \nonumber \\
    {\delta}(\boldsymbol{\mathcal{H}} - \boldsymbol{\mathcal{G}} & {\;\times_{1}\;} \mathbf{A}(\bar{\bm{\phi}} + \bm{\Delta}\bm{\phi}, {\bm{\eta}}, {\mathbf{S}}) \nonumber \\
    & {\;\times_{2}\;} \mathbf{B}(\bar{\bm{\tau}} + \bm{\Delta}\bm{\tau})  {\;\times_{3}\;} \mathbf{C}(\bar{\bm{\nu}} + \bm{\Delta}\bm{\nu}) ).
\end{align}
An interesting perspective is that $\bm{\eta}$ can also be interpreted as perturbation parameters, since they quantify deviations relative to the reference grid at zero (i.e., the far-field case). 

\subsubsection{BDD Domain Channels and SnS Factors}
The low-rank characteristics of SFT domain channels are equivalent to the sparsity of BDD domain channels, which are modeled by Bernoulli-Gaussian distribution, expressed as
\begin{equation}
    \mathsf{P}(\boldsymbol{\mathcal{G}}; \boldsymbol{\mathcal{M}}, \boldsymbol{\mathcal{V}}) {\;\propto\;} \mathsf{BG}(\boldsymbol{\mathcal{G}};  \boldsymbol{\mathcal{M}}, \boldsymbol{\mathcal{C}}(0), \boldsymbol{\mathcal{V}}),
\end{equation}
where $\boldsymbol{\mathcal{M}}$ and $\boldsymbol{\mathcal{V}}$ denote the model hyperparameters, capturing the sparsity and power of BDD domain channels, respectively.

The SnS factor is captured by Bernoulli distributions as
\begin{equation}
    \mathsf{P}(\mathbf{S}; \bm{\Gamma}) {\;\propto\;} \mathsf{exp}(\langle \bm{\Gamma}, \mathbf{S} \rangle),
\end{equation}
where $\bm{\Gamma}{\;\in\;}\mathbb{R}^{N_\text{an}{\times}K_\text{be}}$ denotes the hyperparameters, characterizing the strength of the SnS propagation. 
This model can be extended to structured probabilistic models, such as Markov chains, to capture potential dependencies among SnS factors. It is also applicable to continuous-valued probabilistic models for representing non-binary SnS factors, including Gaussian and Bernoulli-Gaussian distributions. Due to space constraints, this work focuses on the independent binary-valued case; nevertheless, the aforementioned extensions are fully compatible with our framework.

\subsection{Channel Prediction Problem Formulation}\label{sec:EMFramework}
Following these probabilistic models, we formulate the acquisition of BDD domain channels and SnS factors under the MMSE criterion, which can be expressed as the posterior expectations and given by
\begin{equation}
    \hat{\boldsymbol{\mathcal{G}}} = \mathsf{E}\{\boldsymbol{\mathcal{G}}\}, \hat{\mathbf{S}} = \mathsf{E}\{\mathbf{S}\},
\end{equation}
where the expectations are taken with respect to the joint posterior PDF $\mathsf{P}( \boldsymbol{\mathcal{H}}, \boldsymbol{\mathcal{G}}, \mathbf{S} \mid \boldsymbol{\mathcal{Y}}; \mathcal{P}_\text{HP})$, and the hyperparameters are collected as $\mathcal{P}_\text{HP} {\;\triangleq\;} \{\bm{\Delta}\bm{\phi}, \bm{\Delta}\bm{\tau}, \bm{\Delta}\bm{\nu}, {\bm{\eta}},  \boldsymbol{\mathcal{M}}, \boldsymbol{\mathcal{V}}, \bm{\Gamma} \}$. 
Note that the estimation of BDD domain grids is converted into the corresponding hyperparameter learning through the simplification strategy of multilinear transformation models.

However, these posterior expectations require evaluating posterior PDFs that involve unknown hyperparameters, which are environment-dependent and not directly observable in practice.
To address this, we adopt the EM framework to learn these hyperparameters and thereby adapt to diverse propagation environments \cite{moon1996expectation}.
Specifically, the inference iteratively alternates between the E-step and M-step, detailed as follows:
\begin{itemize}
    \item \textbf{E-Step}: Given the current estimate of model hyperparameters, $\hat{\mathcal{P}}_\text{HP}$, from M-step, the posterior PDF $\mathsf{P}( \boldsymbol{\mathcal{H}}, \boldsymbol{\mathcal{G}}, \mathbf{S} \mid \boldsymbol{\mathcal{Y}}; \hat{\mathcal{P}}_\text{HP})$ is computed by combining the observation and multi-linear transformation models, along with the BDD domain channel and SnS factor prior models.
    \item \textbf{M-Step}: Given the posterior PDF, $\mathsf{P}( \boldsymbol{\mathcal{G}} \mid \boldsymbol{\mathcal{Y}}; \hat{\mathcal{P}}_\text{HP})$, derived from the E-step, the model hyperparameters are refined through the maximization of the expected log-likelihood function, expressed as
    \begin{equation}
        \hat{\mathcal{P}}_\text{HP}^{(t_\text{M})} = \arg\max_{\mathcal{P}_\text{HP}} \mathsf{Q}(\mathcal{P}_\text{HP}, \hat{\mathcal{P}}_\text{HP}^{(t_\text{M}-1)}) ,
    \end{equation}
    where $t_\text{M}$ denotes the M-step iteration index, and the Q-function is defined as
    \begin{equation}
        \mathsf{Q}(\mathcal{P}_\text{HP}, \hat{\mathcal{P}}_\text{HP}^{(t_\text{M}-1)}) {\;\triangleq\;} \mathsf{E}\{ \mathsf{ln} (\mathsf{P}( \boldsymbol{\mathcal{Y}}, \boldsymbol{\mathcal{H}}, \boldsymbol{\mathcal{G}}, \mathbf{S}; \mathcal{P}_\text{HP})) \},
    \end{equation}
    where the above expectation is evaluated with respect to $\mathsf{P}( \boldsymbol{\mathcal{H}}, \boldsymbol{\mathcal{G}}, \mathbf{S} \mid \boldsymbol{\mathcal{Y}}; \hat{\mathcal{P}}_\text{HP}^{(t_\text{M}-1)})$ given previous hyperparameter estimations. 
    The M-step iteration index is omitted hereafter to simplify notations. Unless stated otherwise, hyperparameters on the right-hand side of any equation represent values from the previous iteration, while those on the left correspond to the updated ones.
\end{itemize}

\section{Tensor-Structured Bi-Layer Inference for Bayesian Channel Prediction}\label{sec:online_TSDCP}

While the EM framework facilitates MMSE-based channel prediction with unknown hyperparameters, it remains computationally demanding due to the posterior PDF evaluation and hyperparameter learning. 
To alleviate this, we streamline the former in the E-step based on the factor graph principle, and incorporate approximate inference techniques to enhance computational efficiency.
In the M-step, hyperparameters are learned via an alternating optimization strategy to avoid joint search in large parameter spaces, which collectively constitutes the TS-BLI algorithm with the E-step for channel prediction.

\subsection{E-Step: Bi-Layer Factor Graph Representation}\label{sec:inner_layer}

In the E-step, the development of approximate inference techniques is challenged by its underlying affine matrix factorization form: the unknown SnS factors introduce bilinear mixing in the spatial domain, while the frequency and temporal domains preserve the linear mixing of conventional massive MIMO systems.
The unique bilinear structure in the spatial domain leads to extra randomness in the factor matrices beyond the BDD domain channel, which motivates the design of tractable and interpretable Bayesian inference algorithms.

To address this challenge, we introduce the spatial-delay-Doppler (SDD) domain channel to bridge between the SFT and BDD domain channels, allowing only linear mixing between the SDD and SFT domain channels, while bilinear mixing is incorporated between the SDD and BDD domain channels.
This bridge enables a progressive inference trajectory, starting from the SFT domain, advancing through the SDD domain, and ultimately reaching the BDD domain, where each stage involves only one type of mixing: either linear or bilinear.

As such, we comes to the evaluation of the augmented posterior PDF with the SDD domain channel $\boldsymbol{\mathcal{W}} {\;\in\;} \mathbb{C}^{N_\text{an}{\times}K_\text{de}{\times}K_\text{do}}$, which can be factorized by
\begin{align}\label{eq:joint_PDF_factorization}
   \mathsf{P}( \boldsymbol{\mathcal{H}}, \boldsymbol{\mathcal{W}}, & \boldsymbol{\mathcal{G}}, \mathbf{S} \!\mid\! \boldsymbol{\mathcal{Y}}; \hat{\mathcal{P}}_\text{HP}) {\;\propto\;}\mathsf{P}( \boldsymbol{\mathcal{Y}} \!\mid\! \boldsymbol{\mathcal{H}}) \mathsf{P} ( \boldsymbol{\mathcal{H}} \!\mid\! \boldsymbol{\mathcal{W}}; \hat{\bm{\Delta}} \bm{\tau}, \hat{\bm{\Delta}} \bm{\nu} ) \nonumber \\
   & \mathsf{P} ( \boldsymbol{\mathcal{W}} \!\mid\! \boldsymbol{\mathcal{G}}, \mathbf{S}; \hat{\bm{\Delta}} \bm{\phi}, \hat{\bm{\eta}}) \mathsf{P}(\boldsymbol{\mathcal{G}}; \hat{\boldsymbol{\mathcal{M}}}, \hat{\boldsymbol{\mathcal{V}}}) \mathsf{P}({\mathbf{S}}; \hat{\bm{\Gamma}}), 
\end{align}
where the factorized probabilistic models above are denoted, in order, as $\mathsf{P}_{Y}$, $\mathsf{P}_{H}$, $\mathsf{P}_{W}$, $\mathsf{P}_{G}$, and $\mathsf{P}_{S}$, respectively, and the newly introduced probabilistic models are specified by
\begin{subequations}
    \begin{align}
        \mathsf{P} ( \boldsymbol{\mathcal{H}} & \!\mid\! \boldsymbol{\mathcal{W}}; \hat{\bm{\Delta}} \bm{\tau}, \hat{\bm{\Delta}} \bm{\nu} ) {\;\propto\;} \nonumber \\
        &\delta( \boldsymbol{\mathcal{H}} - \boldsymbol{\mathcal{W}} {\;\times_{2}\;} \mathbf{B}(\bar{\bm{\tau}} + \hat{\bm{\Delta}}\bm{\tau}) {\;\times_{3}\;} \mathbf{C}(\bar{\bm{\nu}} + \hat{\bm{\Delta}}\bm{\nu})), \\
        \mathsf{P} ( \boldsymbol{\mathcal{W}} & \!\mid\! \boldsymbol{\mathcal{G}}, \mathbf{S}; \hat{\bm{\Delta}} \bm{\phi}, \hat{\bm{\eta}}) {\;\propto\;} \delta( \boldsymbol{\mathcal{W}} - \boldsymbol{\mathcal{G}} {\;\times_{1}\;} \mathbf{A}(\bar{\bm{\phi}} + \hat{\bm{\Delta}}\bm{\phi}, \hat{\bm{\eta}}, \mathbf{S})).
    \end{align}
\end{subequations}
With these probabilistic models, the equivalence of the MMSE estimators stems from the fact that the posterior PDF required in the E-step can be obtained by marginalizing its augmented counterpart, expressed by
\begin{align}
    \mathsf{P}( \boldsymbol{\mathcal{H}}, \boldsymbol{\mathcal{G}}, & \,\mathbf{S} \mid \boldsymbol{\mathcal{Y}}; \hat{\mathcal{P}}_\text{HP}) {\;\propto\;} \int_{\boldsymbol{\mathcal{W}}} \mathsf{P}( \boldsymbol{\mathcal{H}}, \boldsymbol{\mathcal{W}}, \boldsymbol{\mathcal{G}}, \mathbf{S} \mid \boldsymbol{\mathcal{Y}}; \hat{\mathcal{P}}_\text{HP}).
\end{align}

\begin{figure}[!t]
    \centering
    \includegraphics[width = \linewidth]{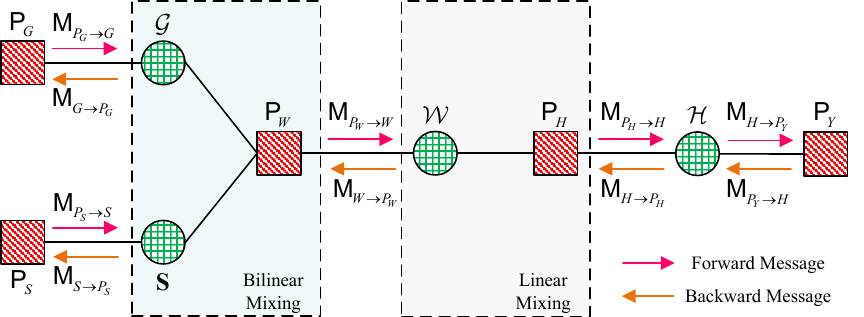}
    \caption{Factor graph representation for E-step, where red squares and green circles represent factor and variable nodes, respectively.}
    \label{fig:FactorGraph}
\end{figure}

Following the factorization of the augmented posterior PDF, the bi-layer factor graph representation employed in the E-step is illustrated in \figref{fig:FactorGraph}, where scalar nodes are merged to improve visual clarity.
The graph is partitioned into two subgraphs corresponding to the linear and bilinear mixing parts, respectively, each leveraging externally defined equivalent prior and likelihood models to facilitate inference \cite{zou2021multi}.
Within this factor graph, we follow the sum-product rules, with messages passed iteratively forward and backward between the observations and BDD domain channels until convergence.
However, exact inference remains computationally intensive due to the high-dimensional integrations involved in posterior expectation computations and the dependencies between messages along the edges.
To mitigate this burden, we employ approximate inference techniques within each layer, based on the central limit theorem and Taylor series, thereby enabling message representations using Gaussian parameterizations and eliminates edge dependencies.

\subsubsection{Linear Mixing Module}
In this module, the message $\mathsf{M}_{P_{W}{\rightarrow}W}$ and $\mathsf{M}_{H{\rightarrow}P_{H}}$ serve as equivalent prior and likelihood models of the linear mixing model, respectively, to develop an equivalent signal model, which is expressed as
\begin{equation}
    \hat{\boldsymbol{\mathcal{H}}}_{\text{lik}} = \boldsymbol{\mathcal{W}} {\;\times_{2}\;} \mathbf{B}(\bar{\bm{\tau}} + \hat{\bm{\Delta}}\bm{\tau}) {\;\times_{3}\;} \mathbf{C}(\bar{\bm{\nu}} + \hat{\bm{\Delta}}\bm{\nu}) + \boldsymbol{\mathcal{Z}}_{H, \text{lik}},
\end{equation}
where $\hat{\boldsymbol{\mathcal{H}}}_{\text{lik}} = \boldsymbol{\mathcal{Y}}$ and $\boldsymbol{\mathcal{Z}}_{H, \text{lik}} = \boldsymbol{\mathcal{Z}}$ denote the equivalent observation and noise of this module, respectively, $\boldsymbol{\mathcal{Z}}_{H, \text{lik}}$ follows an independent Gaussian distribution with zero mean and identical variance ${\sigma}_{z}^{2}$.
By matricizing this model along the frequency and temporal domains, we obtain the linear model with multiple measurement vectors, given by
\begin{equation}
    \hat{\mathbf{H}}_{\text{lik}} = (\mathbf{C}(\bar{\bm{\nu}} + \hat{\bm{\Delta}}\bm{\nu}) {\;\otimes\;} \mathbf{B}(\bar{\bm{\tau}} + \hat{\bm{\Delta}}\bm{\tau}) ) \mathbf{W} + \mathbf{Z}_{H, \text{lik}},
\end{equation}
where we define $\hat{\mathbf{H}}_{\text{lik}}{\;\in\;}\mathbb{C}^{N_\text{sym}N_\text{sc}{\times}N_\text{an}}$, $\mathbf{W}{\;\in\;}\mathbb{C}^{K_\text{do}K_\text{de}{\times}N_\text{an}}$, and $\mathbf{Z}_{H, \text{lik}}{\;\in\;}\mathbb{C}^{N_\text{sym}N_\text{sc}{\times}N_\text{an}}$ as the matrixization of $\hat{\boldsymbol{\mathcal{H}}}_{\text{lik}}$, $\boldsymbol{\mathcal{W}}$, and $\boldsymbol{\mathcal{Z}}_{H, \text{lik}}$ with respect to the frequency and temporal domains, respectively. This implies that the spatial domain observations are treated as independent, allowing this module to focus on the linear mixing part of this multi-linear transformation.

When the number of pilot subcarriers and OFDM symbols tends to infinity, the key input and output messages can be approximated as Gaussian distributions, given by
\begin{subequations}
    \begin{align}
        \mathsf{M}_{P_{H} {\rightarrow} H} &{\;\propto\;} \mathsf{CN}( \boldsymbol{\mathcal{H}}; \hat{\boldsymbol{\mathcal{H}}}_{\text{pri}}, \boldsymbol{\mathcal{E}}_{H, \text{pri}} ),\\
        \mathsf{M}_{W {\rightarrow} P_{W}} &{\;\propto\;} \mathsf{CN}( \boldsymbol{\mathcal{W}}; \hat{\boldsymbol{\mathcal{W}}}_{\text{lik}}, \boldsymbol{\mathcal{E}}_{W, \text{lik}} ),
    \end{align}
\end{subequations}
where the parameters in $\mathsf{M}_{P_{H} {\rightarrow} H}$ and $\mathsf{M}_{W {\rightarrow} P_{W}}$ are computed from Line \ref{state:linear_start} to Line \ref{state:linear_end} of \alref{alg:E_Step}, following the general ideas of the generalized approximate message passing (GAMP) procedure \cite{rangan2011generalized}.
Despite the limited number of pilot OFDM symbols available for channel prediction, the asymptotic Gaussian approximation adopted here remains effective, as confirmed by empirical simulation results.

\subsubsection{Bilinear Mixing Module}
Similar to the linear mixing module, the message $\mathsf{M}_{W \rightarrow P_{W}}$, interpreted as the equivalent likelihood, is combined with the prior models for the BDD domain channels and SnS factors, denoted by $\mathsf{M}_{P_{G} \rightarrow G}$ and $\mathsf{M}_{P_{S} \rightarrow S}$, respectively, to develop the equivalent signal model in this module, given by
\begin{equation}
    \hat{\boldsymbol{\mathcal{W}}}_{\text{lik}} = \boldsymbol{\mathcal{G}} {\;\times_{1}\;} \mathbf{A}(\bar{\bm{\phi}} + \hat{\bm{\Delta}}\bm{\phi}, \hat{\bm{\eta}}, \mathbf{S}) + \boldsymbol{\mathcal{Z}}_{W, \text{lik}},
\end{equation}
where $\boldsymbol{\mathcal{Z}}_{W, \text{lik}}$ denotes the equivalent noise of this module, assumed to follow an independent Gaussian distribution with zero mean and variance $\boldsymbol{\mathcal{E}}_{W, \text{lik}}$. The matrixization along the spatial domain of this model can be expressed as
\begin{equation}
    \hat{\mathbf{W}}_{\text{lik}} = \mathbf{A}(\bar{\bm{\phi}} + \hat{\bm{\Delta}}\bm{\phi}, \hat{\bm{\eta}}, \mathbf{S}) \mathbf{G} + \mathbf{Z}_{W, \text{lik}},
\end{equation}
where we define $\hat{\mathbf{W}}_{\text{lik}}{\;\in\;}\mathbb{C}^{N_\text{an}{\times}K_\text{de}K_\text{do}}$, $\mathbf{G}{\;\in\;}\mathbb{C}^{K_\text{be}{\times}K_\text{de}K_\text{do}}$, and $\mathbf{Z}_{W, \text{lik}}{\;\in\;}\mathbb{C}^{N_\text{an}{\times}K_\text{de}K_\text{do}}$ as the matrixization of $\hat{\boldsymbol{\mathcal{W}}}_{\text{lik}}$, $\boldsymbol{\mathcal{G}}$, and $\boldsymbol{\mathcal{Z}}_{W, \text{lik}}$ with respect to the spatial domain, respectively. 

This module builds upon the structured bilinear formulation, which incorporates the fact that the unknown spatial domain factor matrices is parametrized by SnS factors.
To address such structure, we split it into two sub-modules: matrix factorization (MF) and SnS detection (SnSD), where latter provides the prior model of spatial domain factor matrices and the former returns the likelihood model for the SnS factor detection.

\begin{figure}[!t]
    \centering
    \includegraphics[width = \linewidth]{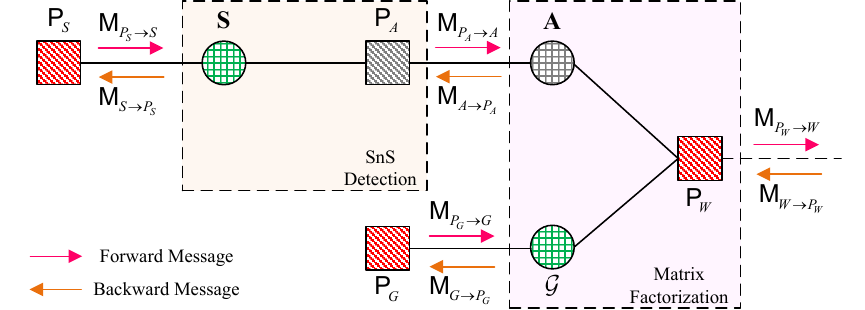}
    \caption{Equivalent factor graph in the bilinear mixing module, where the gray square and circle represent auxiliary factor and variable nodes, respectively.}
    \label{fig:SpatialFactorGraph}
\end{figure}

Specifically, the parameterized structure of spatial domain factor matrices is temporarily disregarded in the MF sub-module, resulting in the simplified bilinear model consistent with the standard formulation. From the factor graph perspective, this can be interpreted as augmenting the model with a new set of auxiliary variable and factor nodes corresponding to the spatial domain factor matrices, as shown in \figref{fig:SpatialFactorGraph}.
When the number of antennas tends to infinity, the Gaussian approximation of input and output messages are given by
\begin{subequations}
    \begin{align}
        \mathsf{M}_{P_{W} {\rightarrow} W} &{\;\propto\;} \mathsf{CN}( \boldsymbol{\mathcal{W}}; \hat{\boldsymbol{\mathcal{W}}}_{\text{pri}}, \boldsymbol{\mathcal{E}}_{W, \text{pri}} ),\\
        \mathsf{M}_{G {\rightarrow} P_{G}} &{\;\propto\;} \mathsf{CN}( \boldsymbol{\mathcal{G}}; \hat{\boldsymbol{\mathcal{G}}}_{\text{lik}}, \boldsymbol{\mathcal{E}}_{G, \text{lik}} ), \\
        \mathsf{M}_{A {\rightarrow} P_{A}} &{\;\propto\;} \mathsf{CN}( \mathbf{A}; \hat{\mathbf{A}}_{\text{lik}}, \boldsymbol{\Sigma}_{A, \text{lik}} ),
    \end{align}
\end{subequations}
where the paramters in $\mathsf{M}_{P_{W} {\rightarrow} W}$, $\mathsf{M}_{G {\rightarrow} P_{G}}$, and $\mathsf{M}_{A {\rightarrow} P_{A}}$ are computed from Line \ref{state:bilinear_start} to Line \ref{state:bilinear_end} of \alref{alg:E_Step}, which follows the bilinear GAMP (BiGAMP) procedure \cite{parker2014bilinear}. 

In the SnSD sub-module, the inference begins with the deterministic parameterized and component-wise structure of spatial domain factor matrices, where the message $\mathsf{M}_{P_{A}{\rightarrow}A}$ is characterized by the Dirac Delta function, given by
\begin{equation}
    \mathsf{M}_{P_{A}{\rightarrow}A} {\;\propto\;} \delta( \mathbf{A} - \mathbf{A}_{\text{SS}} {\;\odot\;} \mathbf{S} ),
\end{equation}
where $\mathbf{A}_{\text{SS}} {\;\in\;} \mathbb{C}^{N_{\text{an}}{\times}K_{\text{be}}}$ denotes the spatial stationary spatial domain factor matrix with the hyperparameters $\hat{\bm{\Delta}}\bm{\phi}$ and $\hat{\bm{\eta}}$ from M-step, defnied by
\begin{equation}
    \mathbf{A}_{\text{SS}} = [ \mathbf{a}_{\text{SS}}(\bar{\phi}_{1}\!+\!\hat{{\Delta}}{\phi}_{1}, \hat{\eta}_{1}), {\dots}, \mathbf{a}_{\text{SS}}(\bar{\phi}_{K_\text{be}}\!+\!\hat{{\Delta}}{\phi}_{K_\text{be}}, \hat{\eta}_{K_\text{be}}) ].
\end{equation}

\subsubsection{Closing the Loop}
To close the loop, we start with the missing messages in this bi-layer factor graph. Specifically, the messages $\mathsf{M}_{P_{G}{\rightarrow}G}$, $\mathsf{M}_{P_{S}{\rightarrow}S}$, and $\mathsf{M}_{P_{Y}{\rightarrow}H}$ are determined by the BDD domain channel prior model, the SnS factor prior model, and the observation model, respectively. 

The posterior estimation of hidden variables involved in messages also plays a critical role. Following the sum-product rules, the posterior PDFs of SFT, SDD, and BDD domain channels can be directly obtained by
\begin{subequations}
    \begin{align}
        \mathsf{b}_{H} &{\;\propto\;} \mathsf{M}_{P_{H}{\rightarrow}H} \mathsf{M}_{H{\rightarrow}P_{H}}, \label{eq:postPDF_H} \\
        \mathsf{b}_{W} &{\;\propto\;} \mathsf{M}_{P_{W}{\rightarrow}W} \mathsf{M}_{W{\rightarrow}P_{W}}, \label{eq:postPDF_W} \\
        \mathsf{b}_{G} &{\;\propto\;} \mathsf{M}_{P_{G}{\rightarrow}G} \mathsf{M}_{G{\rightarrow}P_{G}}, \label{eq:postPDF_G}
    \end{align}
\end{subequations}
where we have $\mathsf{M}_{H{\rightarrow}P_{H}} {\;\propto\;} \mathsf{M}_{P_{Y}{\rightarrow}H}$, and the posterior estimates are computed by evaluating component-wise expectations over the corresponding PDFs.

For spatial domain factor matrices, the equivalent likelihood model of SnS factors is specified by the message $\mathsf{M}_{A {\rightarrow} P_{A}}$ from the MF sub-module and channel model, expressed as
\begin{equation}
    \hat{\mathbf{A}}_{\text{lik}} =  \mathbf{A}_{\text{SS}} {\;\odot\;} \mathbf{S} + \mathbf{Z}_{A, \text{lik}},
\end{equation}
where $\mathbf{Z}_{A, \text{lik}}$ denotes the equivalent noise with variance $\boldsymbol{\Sigma}_{A, \text{lik}}$. Therefore, the posterior PDF is given by
\begin{equation}\label{eq:posteriorPDF_SDFM}
    \mathsf{P}(\mathbf{S} \!\mid\! \hat{\mathbf{A}}_{\text{lik}}; \hat{\bm{\Gamma}}) {\;\propto\;} \mathsf{exp}(\langle \hat{\bm{\Gamma}}, \mathbf{S} \rangle) \mathsf{CN}(\hat{\mathbf{A}}_{\text{lik}}; \mathbf{A}_{\text{SS}} {\;\odot\;} \mathbf{S}, \boldsymbol{\Sigma}_{A, \text{lik}}),
\end{equation}
Using the above posterior PDF, the posterior mean and variance of spatial domain factor matrices can be given by
\begin{subequations}\label{eq:posterior_SDFM}
    \begin{equation}\label{eq:posterior_SDFM_mean}
        \hat{\mathbf{A}} = \mathbf{A}_{\text{SS}} {\;\odot\;} \hat{\mathbf{S}},
    \end{equation}
    \begin{equation}\label{eq:posterior_SDFM_var}
        \bm{\Sigma}_{A, \text{post}} = |\mathbf{A}_{\text{SS}}|^{{\odot}2} {\;\odot\;} \hat{\mathbf{S}} {\;\odot\;} (1 - \hat{\mathbf{S}}),
    \end{equation}
\end{subequations}
where $\hat{\mathbf{S}}$ denotes the posterior expectation of $\mathbf{S}$.

\begin{algorithm}[!t]
  \caption{Single Iteration of E-Step}
  \label{alg:E_Step}
  \begin{algorithmic}[1]
  \Require {External Information ($\boldsymbol{\mathcal{Y}}, {\sigma}_{z}^{2}$), hyperparameters estimated from M-step ($\hat{\mathcal{P}}_{\text{HP}}$), and intermediate results from the previous E-step.}
  \Ensure {Intermediate results to the next E- and M-step.}
  \For {$t_\text{E} = 1, {\dots}, T_\text{E}$}
  \Statex {\textbf{// Linear Mixing Module}}
  \State {$\boldsymbol{\mathcal{E}}_{H, \text{pri}} = \boldsymbol{\mathcal{E}}_{W, \text{post}} {\;\times_{2}\;} |\mathbf{B}|^{{\odot}2}  {\;\times_{3}\;} |\mathbf{C}|^{{\odot}2} $}\label{state:linear_start}
  \State {$\hat{\boldsymbol{\mathcal{H}}}_{\text{pri}} = \hat{\boldsymbol{\mathcal{W}}} {\;\times_{2}\;} \mathbf{B} {\;\times_{3}\;} \mathbf{C} - \hat{\boldsymbol{\mathcal{H}}}_{\text{res}} {\;\odot\;} \boldsymbol{\mathcal{E}}_{H, \text{pri}} $}
  \State {Obtain the posterior mean $\hat{\boldsymbol{\mathcal{H}}}$ and variance $\boldsymbol{\mathcal{E}}_{H, \text{post}}$ of $\boldsymbol{\mathcal{H}}$ based on \eqref{eq:postPDF_H}.}
  \State {$\boldsymbol{\mathcal{E}}_{H, \text{res}} = (\boldsymbol{\mathcal{E}}_{H, \text{post}} - \boldsymbol{\mathcal{E}}_{H, \text{pri}}) {\;\oslash\;} (\boldsymbol{\mathcal{E}}_{H, \text{pri}})^{{\odot}2} $}
  \State {$\hat{\boldsymbol{\mathcal{H}}}_{\text{res}} = (\hat{\boldsymbol{\mathcal{H}}} - \hat{\boldsymbol{\mathcal{H}}}_{\text{pri}}) {\;\oslash\;} \boldsymbol{\mathcal{E}}_{H, \text{pri}} $}
  \State {$\boldsymbol{\mathcal{E}}_{W, \text{lik}} = (\boldsymbol{\mathcal{E}}_{H, \text{res}} {\;\times_{2}\;} |\mathbf{B}^{H}|^{{\odot}2} {\;\times_{3}\;} |\mathbf{C}^{H}|^{{\odot}2} )^{{\odot}-1} $}
  \State {$\hat{\boldsymbol{\mathcal{W}}}_{\text{lik}} = \hat{\boldsymbol{\mathcal{W}}} + \boldsymbol{\mathcal{E}}_{W, \text{lik}} {\;\odot\;}(\boldsymbol{\mathcal{H}}_{\text{res}} {\;\times_{2}\;} \mathbf{B}^{H} {\;\times_{3}\;} \mathbf{C}^{H}) $}\label{state:linear_end}
  \Statex {\textbf{// Bilinear Mixing Module}}
  \State {$\boldsymbol{\mathcal{E}}_{W, \text{plug-in}} = \boldsymbol{\mathcal{E}}_{G, \text{post}} {\;\times_{1}\;} |\hat{\mathbf{A}}|^{{\odot}2} + |\hat{\boldsymbol{\mathcal{G}}}|^{{\odot}2} {\;\times_{1}\;} \bm{\Sigma}_{A, \text{post}} $}\label{state:bilinear_start}
  \State {$\hat{\boldsymbol{\mathcal{W}}}_{\text{pri}} = \hat{\boldsymbol{\mathcal{G}}} {\;\times_{1}\;} \hat{\mathbf{A}} - \hat{\boldsymbol{\mathcal{W}}}_{\text{res}} {\;\odot\;} \boldsymbol{\mathcal{E}}_{W, \text{plug-in}} $}
  \State {$\boldsymbol{\mathcal{E}}_{W, \text{pri}} = \boldsymbol{\mathcal{E}}_{W, \text{plug-in}} + \boldsymbol{\mathcal{E}}_{G, \text{pri}} {\;\times_{1}\;} \bm{\Sigma}_{A, \text{post}} $}
  \State {Obtain the posterior mean $\hat{\boldsymbol{\mathcal{W}}}$ and variance $\boldsymbol{\mathcal{E}}_{W, \text{post}}$ of $\boldsymbol{\mathcal{W}}$ based on \eqref{eq:postPDF_W}.}
  \State {$\boldsymbol{\mathcal{E}}_{W, \text{res}} = (\boldsymbol{\mathcal{E}}_{W, \text{post}} - \boldsymbol{\mathcal{E}}_{W, \text{pri}}) {\;\oslash\;} (\boldsymbol{\mathcal{E}}_{W, \text{pri}})^{{\odot}2} $}
  \State {$\hat{\boldsymbol{\mathcal{W}}}_{\text{res}} = (\hat{\boldsymbol{\mathcal{W}}} - \hat{\boldsymbol{\mathcal{W}}}_{\text{pri}}) {\;\oslash\;} \boldsymbol{\mathcal{E}}_{W, \text{pri}} $}
  \State {$\boldsymbol{\mathcal{E}}_{G, \text{lik}} = (\boldsymbol{\mathcal{E}}_{W, \text{res}} {\;\times_{1}\;} |\hat{\mathbf{A}}^{H}|^{{\odot}2} )^{{\odot}-1} $}
  \State \parbox[t]{\dimexpr\linewidth-\algorithmicindent\relax}{
      $\hat{\boldsymbol{\mathcal{G}}}_{\text{lik}} = \hat{\boldsymbol{\mathcal{G}}} - \hat{\boldsymbol{\mathcal{G}}} \odot \boldsymbol{\mathcal{E}}_{G, \text{lik}} \odot ( \boldsymbol{\mathcal{E}}_{W, \text{res}} \times_{1} \bm{\Sigma}_{A, \text{post}} )$\\
      \hspace*{2.5em} $\, + \, \boldsymbol{\mathcal{E}}_{G, \text{lik}} \odot (\hat{\boldsymbol{\mathcal{W}}}_{\text{res}} \times_{1} \hat{\mathbf{A}}^{H} )$
  }
  \State {Obtain the posterior mean $\hat{\boldsymbol{\mathcal{G}}}$ and variance $\boldsymbol{\mathcal{E}}_{G, \text{post}}$ of $\boldsymbol{\mathcal{G}}$ based on \eqref{eq:postPDF_G}.}
  \State {$\bm{\Sigma}_{A, \text{lik}} = (\boldsymbol{\mathcal{E}}_{W, \text{res}} {\;\times_{-1}\;} |\hat{\boldsymbol{\mathcal{G}}}|^{{\odot}2} )^{{\odot}-1}$}
  \State \parbox[t]{\dimexpr\linewidth-\algorithmicindent\relax}{
      $\hat{\mathbf{A}}_{\text{lik}} = \hat{\mathbf{A}} - \hat{\mathbf{A}} {\;\odot\;} \bm{\Sigma}_{A, \text{lik}} {\;\odot\;} (\boldsymbol{\mathcal{E}}_{W, \text{res}} {\;\times_{-1}\;} \boldsymbol{\mathcal{E}}_{G, \text{post}} )$\\
      \hspace*{2.5em} $\, + \, \bm{\Sigma}_{A, \text{lik}} \odot (\hat{\boldsymbol{\mathcal{W}}}_{\text{res}} \times_{-1} \hat{\boldsymbol{\mathcal{G}}}^{\ast} )$
  }\label{state:bilinear_end}
  \State {Obtain the posterior mean $\hat{\mathbf{A}}$ and variance $\bm{\Sigma}_{A, \text{post}}$ of $\mathbf{A}$ based on \eqref{eq:posteriorPDF_SDFM} and \eqref{eq:posterior_SDFM}.}
  \EndFor
  \end{algorithmic}
\end{algorithm}

In summary, the message passing process between these two modules is presented in \alref{alg:E_Step} as one E-step, where $\hat{\mathbf{A}} {\;\triangleq\;} \mathbf{A}(\bar{\bm{\phi}} + \hat{\bm{\Delta}}\bm{\phi}, \hat{\bm{\eta}}, \hat{\mathbf{S}})$, $\mathbf{B}{\;\triangleq\;}\mathbf{B}(\bar{\bm{\tau}} + \hat{\bm{\Delta}}\bm{\tau})$, and $\mathbf{C}{\;\triangleq\;}\mathbf{C}(\bar{\bm{\nu}} + \hat{\bm{\Delta}}\bm{\nu})$ denote shorthand representations of factor matrices under estimated SnS factors and hyperparameters.

\subsection{M-Step: Model Hyperparameter Learning}\label{sec:outer_layer}
In the M-step, we employ an alternating optimization strategy to reduce the hyperparameter search space. Specifically, we start with perturbation and slope parameters in the factor matrices, with updating rules decoupled as
\begin{subequations}\label{eq:original_learning_rules_ml}
    \begin{align}
        \{ \hat{\bm{\Delta}}\bm{\tau}, \hat{\bm{\Delta}}\bm{\nu} \} &= \mathop{\arg\max}_{\bm{\Delta}\bm{\tau}, \bm{\Delta}\bm{\nu}} \underbrace{\mathsf{E}\{ \mathsf{ln} ( \mathsf{P} ( \boldsymbol{\mathcal{H}} \!\mid\! \boldsymbol{\mathcal{W}}; {\bm{\Delta}} \bm{\tau}, {\bm{\Delta}} \bm{\nu} )) \}}_{{\;\triangleq\;}-J_{\tau, \nu}(\bm{\Delta}\bm{\tau}, \bm{\Delta}\bm{\nu})}, \label{eq:original_learning_rules_linear} \\
        \{ \hat{\bm{\Delta}}\bm{\phi}, \hat{\bm{\eta}} \} &= \mathop{\arg\max}_{\bm{\Delta}\bm{\phi}, {\bm{\eta}}} \underbrace{\mathsf{E}\{ \mathsf{ln} ( \mathsf{P} ( \boldsymbol{\mathcal{W}} \!\mid\! \boldsymbol{\mathcal{G}}, \mathbf{S}; \bm{\Delta}\bm{\phi}, {\bm{\eta}} )) \}}_{{\;\triangleq\;}-J_{\phi, \eta}(\bm{\Delta}\bm{\phi}, {\bm{\eta}})}, \label{eq:original_learning_rules_bilinear}
    \end{align}
\end{subequations}
where the expectation is taken over the posterior PDFs from the E-step, and such decoupling inherently results from the factorized forms in \eqref{eq:joint_PDF_factorization}.
The objective functions can be further simplified as the negative residual energy of SFT and SDD domain channels, as summarized in \propref{prop:bdd_objectivefunc}.

\begin{proposition}\label{prop:bdd_objectivefunc}
    The negative objective functions $J_{\tau, \nu}$ and $J_{\phi, \eta}$ are given by
    \begin{subequations}\label{eq:bdd_objectivefunc}
        \begin{gather}
            J_{\tau, \nu} \!=\! \|  \hat{\boldsymbol{\mathcal{H}}} - \hat{\boldsymbol{\mathcal{W}}} {\;\times_{2}\;} \mathbf{B}(\bar{\bm{\tau}} + \bm{\Delta}\bm{\tau}) {\;\times_{3}\;} \mathbf{C}(\bar{\bm{\nu}} + \bm{\Delta}\bm{\nu}) \|_{F}^{2}, \label{eq:linear_objectivefunc} \\ 
            J_{\phi, \eta} \!=\! \|  \hat{\boldsymbol{\mathcal{W}}} - \hat{\boldsymbol{\mathcal{G}}}{\;\times_{1}\;} \mathbf{A}(\bar{\bm{\phi}} + \bm{\Delta}\bm{\phi}, {\bm{\eta}}, \hat{\mathbf{S}}) \|_{F}^{2}, \label{eq:bilinear_objectivefunc}
        \end{gather}
    \end{subequations}
    respectively, where $\hat{\boldsymbol{\mathcal{H}}}$, $\hat{\boldsymbol{\mathcal{W}}}$, $\hat{\boldsymbol{\mathcal{G}}}$, and $\hat{\mathbf{S}}$ denote the posterior estimates obtained from the previous E-step.
    \begin{proof}
        By approximating the Dirac Delta function in \eqref{eq:original_learning_rules_linear} with the sharply peaked Gaussian PDF, the objective function of linear mixing is given by \cite{hou2024tensor}
        \begin{equation}
            J_{\tau, \nu} = \mathsf{E}\{ \|  \boldsymbol{\mathcal{H}} - \boldsymbol{\mathcal{W}} {\;\times_{2}\;} \mathbf{B}(\bar{\bm{\tau}} + \bm{\Delta}\bm{\tau}) {\;\times_{3}\;} \mathbf{C}(\bar{\bm{\nu}} + \bm{\Delta}\bm{\nu}) \|_{F}^{2} \},
        \end{equation}
        With the posterior independence derived from the E-step, the objective function is further decomposed into $J_{\tau, \nu} = J_{\tau, \nu, 1} + J_{\tau, \nu, 2}$, where sub-functions are defined as
        \begin{subequations}
            \begin{equation}
                J_{\tau, \nu, 1} {\;\triangleq\;}  \|  \hat{\boldsymbol{\mathcal{H}}} - \hat{\boldsymbol{\mathcal{W}}} {\;\times_{2}\;} \mathbf{B}(\bar{\bm{\tau}} + \bm{\Delta}\bm{\tau}) {\;\times_{3}\;} \mathbf{C}(\bar{\bm{\nu}} + \bm{\Delta}\bm{\nu}) \|_{F}^{2},
            \end{equation}
            \begin{equation}
                J_{\tau, \nu, 2} {\;\triangleq\;} \| \boldsymbol{\mathcal{E}}_{W, \text{post}} {\;\times_{2}\;} |\mathbf{B}(\bar{\bm{\tau}} + \bm{\Delta}\bm{\tau})|^{{\odot}2} {\;\times_{3}\;} |\mathbf{C}(\bar{\bm{\nu}} + \bm{\Delta}\bm{\nu})|^{{\odot}2} \|_{1}.
            \end{equation}
        \end{subequations}
        Since the magnitudes of the factor matrices are, by definition, independent of $\bm{\Delta}\bm{\tau}$ and $\bm{\Delta}\bm{\nu}$, the sub-function $J_{\tau, \nu, 2}$ is also constant with respect to these variables. Therefore, the objective function for hyperparameter learning corresponding to linear mixing reduces to \eqref{eq:linear_objectivefunc}, and the bilinear mixing case can be derived analogously as \eqref{eq:bilinear_objectivefunc}.
    \end{proof}
\end{proposition}

By leveraging the first-order Taylor series of factor matrices, these objective functions can be further approximated as quadratic forms and summarized in \propref{prop:bdd_hp}.

\begin{proposition}\label{prop:bdd_hp}
    The negative objective functions in \eqref{eq:bdd_objectivefunc} are approximated as the following quadratic forms, given by
    \begin{subequations}
        \begin{align}
            J_{\phi,  \eta}(\bm{\chi}) &{\;\approx\;} \bm{\chi}^{T} \bm{\Pi}_{\phi,  \eta} \bm{\chi} - 2 \mathsf{Re}\{\bm{\mu}_{\phi,  \eta}^{T}\} \bm{\chi} + C_{\phi,  \eta}, \\
            J_{\tau}(\bm{\Delta}\bm{\tau}) &{\;\approx\;} \bm{\Delta}\bm{\tau}^{T} \bm{\Pi}_{\tau} \bm{\Delta}\bm{\tau} - 2 \mathsf{Re}\{\bm{\mu}_{\tau}^{T}\} \bm{\Delta}\bm{\tau} + C_{\tau}, \\
            J_{\nu}(\bm{\Delta}\bm{\nu}) &{\;\approx\;} \bm{\Delta}\bm{\nu}^{T} \bm{\Pi}_{\nu} \bm{\Delta}\bm{\nu} - 2 \mathsf{Re}\{ \bm{\mu}_{\nu}^{T} \} \bm{\Delta}\bm{\nu} + C_{\nu},
        \end{align}
    \end{subequations}
    where $\bm{\chi} {\;\triangleq\;} [\bm{\Delta}\bm{\phi}^{T}, \bm{\eta}^{T}]^{T}$ denotes beam domain perturbation and slope parameters, $C_{\phi,  \eta}$, $C_{\tau}$, and $C_{\nu}$ denote constant terms, the quadratic and linear coefficients are given by
    \begin{subequations}
        \begin{gather}
            \bm{\Pi}_{\phi, \eta} = (\dot{\mathbf{A}}^{H}(\bar{\bm{\phi}}, \bm{0}, \hat{\mathbf{S}})\dot{\mathbf{A}}(\bar{\bm{\phi}}, \bm{0}, \hat{\mathbf{S}}))^{\ast} {\;\odot\;} ( \bm{1}_{2{\times}2} {\;\otimes\;} (\hat{\boldsymbol{\mathcal{G}}} {\;\times_{-1}\;} \hat{\boldsymbol{\mathcal{G}}}^{\ast}) ), \\
            \bm{\mu}_{\phi,  \eta} = \sum_{n} \mathsf{diag}^{H}\{ \bm{1}_{2{\times}1} {\;\otimes\;} \hat{\mathbf{g}}_{n} \} \dot{\mathbf{A}}^{H}(\bar{\bm{\phi}}, \bm{0}, \hat{\mathbf{S}}) \hat{\mathbf{r}}_{\phi,  \eta, n}, \\
            \bm{\Pi}_{\tau} = (\dot{\mathbf{B}}^{H} (\bar{\bm{\tau}}) \dot{\mathbf{B}}(\bar{\bm{\tau}}))^{\ast} {\;\odot\;} (\hat{\boldsymbol{\mathcal{W}}}_{\tau} {\;\times_{-2}\;} \hat{\boldsymbol{\mathcal{W}}}_{\tau}^{\ast}), \\
            \bm{\mu}_{\tau} = \sum_{n} \mathsf{diag}^{H}\{ \hat{\mathbf{w}}_{\tau, n} \} \dot{\mathbf{B}}^{H} (\bar{\bm{\tau}}) \hat{\mathbf{r}}_{\tau, n} , \\
            \bm{\Pi}_{\nu} = (\dot{\mathbf{C}}^{H} (\bar{\bm{\nu}}) \dot{\mathbf{C}}(\bar{\bm{\nu}}))^{\ast} {\;\odot\;} (\hat{\boldsymbol{\mathcal{W}}}_{\nu} {\;\times_{-3}\;} \hat{\boldsymbol{\mathcal{W}}}_{\nu}^{\ast}), \\
            \bm{\mu}_{\nu} = \sum_{n} \mathsf{diag}^{H}\{ \hat{\mathbf{w}}_{\nu, n} \} \dot{\mathbf{C}}^{H} (\bar{\bm{\nu}}) \hat{\mathbf{r}}_{\nu, n} ,
        \end{gather}
    \end{subequations}
    where $\dot{\mathbf{A}}(\cdot, \cdot, \cdot) {\;\triangleq\;} [\dot{\mathbf{A}}_{\phi}(\cdot, \cdot, \cdot), \dot{\mathbf{A}}_{\eta}(\cdot, \cdot, \cdot)]$ denotes the aggregate derivative matrix, $\dot{\mathbf{A}}_{\phi}(\cdot, \cdot, \cdot)$ and $\dot{\mathbf{A}}_{\eta}(\cdot, \cdot, \cdot)$ denote the first-order derivatives of $\mathbf{A}(\cdot, \cdot, \cdot)$ with respect to the first and second arguments, respectively,  $\dot{\mathbf{B}}(\cdot)$ and $\dot{\mathbf{C}}(\cdot)$ denote the first-order derivatives of $\mathbf{B}(\cdot)$ and $\mathbf{C}(\cdot)$, respectively, $\hat{\mathbf{g}}_{n}$, $\hat{\mathbf{w}}_{\tau, n}$ and $\hat{\mathbf{w}}_{\nu, n}$ denotes the $n$-th fibers of beam-delay-Doppler domain channel $\hat{\boldsymbol{\mathcal{G}}}$, spatial-delay-temporal domain channel $\hat{\boldsymbol{\mathcal{W}}}_{\tau} {\;\triangleq\;} \hat{\boldsymbol{\mathcal{W}}} {\;\times_{3}\;} \mathbf{C} (\bar{\bm{\nu}} + \hat{\bm{\Delta}}\bm{\nu})$ and spatial-frequency-Doppler domain channel $\hat{\boldsymbol{\mathcal{W}}}_{\nu}{\;\triangleq\;} \hat{\boldsymbol{\mathcal{W}}} {\;\times_{2}\;}\mathbf{B} (\bar{\bm{\tau}} + \hat{\bm{\Delta}}\bm{\tau})$, respectively, $\hat{\mathbf{r}}_{\phi,  \eta, n}$, $\hat{\mathbf{r}}_{\tau, n}$, and $\hat{\mathbf{r}}_{\nu, n}$ denote the $n$-th fibers of residual channels $\hat{\boldsymbol{\mathcal{R}}}_{\phi,  \eta}$, $\hat{\boldsymbol{\mathcal{R}}}_{\tau}$, and $\hat{\boldsymbol{\mathcal{R}}}_{\nu}$ along spatial, frequency, and temporal domains, respectively, with corresponding residual channels defined as
    \begin{subequations}
        \begin{gather}
            \hat{\boldsymbol{\mathcal{R}}}_{\phi,  \eta} = \hat{\boldsymbol{\mathcal{W}}} - \hat{\boldsymbol{\mathcal{G}}} {\;\times_{1}\;} \mathbf{A}(\bar{\bm{\phi}}, \bm{0}, \hat{\mathbf{S}}), \\
            \hat{\boldsymbol{\mathcal{R}}}_{\tau} = \hat{\boldsymbol{\mathcal{H}}} - \hat{\boldsymbol{\mathcal{W}}} {\;\times_{2}\;} \mathbf{B} (\bar{\bm{\tau}}) {\;\times_{3}\;} \mathbf{C} (\bar{\bm{\nu}} + \hat{\bm{\Delta}}\bm{\nu}), \\
            \hat{\boldsymbol{\mathcal{R}}}_{\nu} = \hat{\boldsymbol{\mathcal{H}}} - \hat{\boldsymbol{\mathcal{W}}} {\;\times_{2}\;} \mathbf{B} (\bar{\bm{\tau}} + \hat{\bm{\Delta}}\bm{\tau}) {\;\times_{3}\;} \mathbf{C} (\bar{\bm{\nu}}).
        \end{gather}
    \end{subequations}
    \begin{proof}
        The proof follows a similar approach to our previous work \cite[Appendix~C]{hou2024tensor}.
    \end{proof}
\end{proposition}
Based on the approximation of objective functions, closed-form learning rules for the hyperparameters can be derived. Alternatively, the Taylor series can be expanded around the previous learning result, enabling the reuse of intermediate tensors in E-step to avoid redundant computations.

\begin{figure*}[!t]
  \normalsize
  \begin{equation}\label{eq:LR_G}
      [\boldsymbol{\mathcal{R}}_{G}]_{k_\text{be}, k_\text{de}, k_\text{do}} = \frac{[\hat{\boldsymbol{\mathcal{M}}}]_{k_\text{be}, k_\text{de}, k_\text{do}} \mathsf{CN}([\hat{\boldsymbol{\mathcal{G}}}_{\text{lik}}]_{k_\text{be}, k_\text{de}, k_\text{do}}; 0, [\hat{\boldsymbol{\mathcal{V}}}]_{k_\text{be}, k_\text{de}, k_\text{do}} + [\boldsymbol{\mathcal{E}}_{G, \text{lik}}]_{k_\text{be}, k_\text{de}, k_\text{do}})}{(1 - [\hat{\boldsymbol{\mathcal{M}}}]_{k_\text{be}, k_\text{de}, k_\text{do}}) \mathsf{CN}([\hat{\boldsymbol{\mathcal{G}}}_{\text{lik}}]_{k_\text{be}, k_\text{de}, k_\text{do}}; 0, [\boldsymbol{\mathcal{E}}_{G, \text{lik}}]_{k_\text{be}, k_\text{de}, k_\text{do}})}.
  \end{equation}
  \hrulefill
\end{figure*}

When it comes to hyperparameters in the SnS factor prior model, we have
\begin{align}
    \hat{\bm{\Gamma}} = \arg \max_{\bm{\Gamma}} \mathsf{E}\{\mathsf{ln} \mathsf{P}(\mathbf{S} ; \bm{\Gamma} )\}.
\end{align}
By taking the derivative of $\bm{\Gamma}$ and setting it to zero, the learning rules are given by
\begin{equation}\label{eq:updating_rule_SnS}
    \hat{\bm{\Gamma}} = \hat{\mathbf{S}}{\;\oslash\;}(1 + \hat{\mathbf{S}}).
\end{equation}
Since the hyperparamter learning rules in the BDD domain prior channel model are consistent with the techniques in \cite{vila2013expectation}, the detailed derivations are omitted due to the space limitations, with the results summarized as
\begin{subequations}\label{eq:updating_rule_BDD}
    \begin{equation}
        \hat{\boldsymbol{\mathcal{M}}} = \boldsymbol{\mathcal{R}}_{G} {\;\oslash\;} (\boldsymbol{\mathcal{C}}(1) + \boldsymbol{\mathcal{R}}_{G}),
    \end{equation}
    \begin{equation}
        \hat{\boldsymbol{\mathcal{V}}} = ( \boldsymbol{\mathcal{E}}_{G, \text{post}} + |\hat{\boldsymbol{\mathcal{G}}}|^{{\odot}2} ) {\;\oslash\;} \hat{\boldsymbol{\mathcal{M}}},
    \end{equation}
\end{subequations}
where $\boldsymbol{\mathcal{R}}_{G}$ is defined in \eqref{eq:LR_G} at the top of the next page.

\subsection{Bayesian Channel Prediction}

\begin{algorithm}[!t]
  \caption{Tensor-Structured Bi-Layer Inference}
  \label{alg:TSBCP}
  \begin{algorithmic}[1]
  \Require {Observation tensor $\boldsymbol{\mathcal{Y}}$ and noise variance ${\sigma}_{z}^{2}$.}
  \Ensure {Channel Prediction result $\hat{\boldsymbol{\mathcal{H}}}^{\text{CP}}$.}
  \State {Initialize $\hat{\boldsymbol{\mathcal{M}}}$, $\hat{\boldsymbol{\mathcal{V}}}$, $\hat{\bm{\Gamma}}$, $\hat{\boldsymbol{\mathcal{G}}}$, $\hat{\mathbf{S}}$, and all hyperparameters.}
  \For {$t_{\text{M}} = 1, {\dots}, T_\text{M}$}
  \State {Execute Algorithm 1.}
  \State {Learn hyperparameters based on \propref{prop:bdd_objectivefunc}, \propref{prop:bdd_hp}, \eqref{eq:updating_rule_SnS}, and \eqref{eq:updating_rule_BDD}.}
  \EndFor
  \State {Predict the channel based on \eqref{eq:channel_prediction}.}
  \end{algorithmic}
\end{algorithm}

The proposed tensor-structured bi-layer inference algorithm for channel prediction is summarized as \alref{alg:TSBCP}, where posterior estimators with damping are employed in the E-step to guarantee the convergence.
The channel prediction is achieved by the transformation from Doppler domain to temporal domain, formulated as
\begin{align}\label{eq:channel_prediction}
    \boldsymbol{\mathcal{H}}^{\text{CP}} = \hat{\boldsymbol{\mathcal{G}}} &{\;\times_{1}\;} \mathbf{A}(\bar{\bm{\phi}} + \hat{\bm{\Delta}}\bm{\phi}, \hat{\bm{\eta}}, \hat{\mathbf{S}}) \nonumber \\
    &{\;\times_{2}\;} \mathbf{B}(\bar{\bm{\tau}} + \hat{\bm{\Delta}}{\bm{\tau}}) {\;\times_{3}\;} \tilde{\mathbf{C}}(\bar{\bm{\nu}} + \hat{\bm{\Delta}}{\bm{\nu}}), 
\end{align}
where $\tilde{\mathbf{C}}(\bm{\nu}) = [ \tilde{\mathbf{c}}({\nu}_{1}), {\dots}, \tilde{\mathbf{c}}({\nu}_{K_\text{do}}) ]$ denotes the factor matrix in the temporal domain for channel prediction, $\tilde{\mathbf{c}}(\nu)$ denotes the Doppler domain steering vector for channel prediction, defined as $[\tilde{\mathbf{c}}(\nu)]_{n_\text{cp}} = \mathsf{exp}(j2{\pi}(T_{0} + n_\text{cp}{\Delta}T){\nu})$, $T_{0} = (N_\text{sym}-1){\Delta}\bar{T}$ and $n_\text{cp}$ denote the prediction origin and prediction length, respectively.

The prefix “tensor-structured” of \alref{alg:TSBCP} stems from the inherent multi-linear structure of channels, which allows for the natural decoupling of the linear and bilinear mixing, as well as the simplification of hyperparameter learning in the multi-linear transformations. This structure not only supports bi-layer inference across different types of mixing, but also improves computational efficiency via tensor operations, as further analyzed in \secref{sec:complexity}.

\begin{remark}
    An alternative interpretation of EM arises from the variational free energy perspective \cite{zhang2021unifying}, where it is regarded as a special case of message passing under an uninformative Dirac delta prior. While not explicitly employed in this work, it offers useful intuition for the message schedule design. Specifically, the E-step need not fully converge before proceeding to the M-step, thereby alleviating the computational burden when hyperparameters remain unreliable in early iterations.
\end{remark}

\section{Simulation Results}\label{sec:simulation_result}
\subsection{Simulation Configuration}

\begin{table}[!t]
  \caption{Scenario Parameters}
  \label{tab:simulation_configuration}
  \centering
  \begin{tabular}{cc}
  \toprule
  \textbf{Paramter} & \textbf{Value} \\
  \midrule
  Carrier Frequency & $f_{\text{c}} = 15$ GHz \\
  OFDM Symbol Duration & ${\Delta}T_{\text{sym}} = 16.67$ $\mu$s \\
  Cyclic Prefix Duration & ${\Delta}T_{\text{cp}} = 1.17$ $\mu$s \\
  Subcarrier Spacing & ${\Delta}f = 60$ kHz \\
  Number of Pilot Interval Symbols & $N_\text{IS} = 14$ \\
  Number of Transmission Combs & $N_\text{TC} = 4$ \\
  Number of BS Antennas & $N_{\text{an}} = 128$ \\
  Number of Pilot Subcarriers & $N_\text{sc} = 128$ \\
  Number of Pilot Symbols & $ N_\text{sym} = 10 $ \\
  Minimum MT Distribution Radius & $r_{\text{min}} = 10$ m \\
  Velocity of MTs & $v_\text{MT} = 60$ km/h \\
  \bottomrule
  \end{tabular}
\end{table}

\subsubsection{Scenario Setting}
To validate the exactness of proposed channel and probabilistic models, we employ the QuaDRiGa channel simulator, which generates XL-MIMO-OFDM channels consistent with the third Generation Partnership Program (3GPP) specifications \cite{3gpp38901} and has been validated in various field trials \cite{jaeckel2014quadriga}. Specifically, we consider the 3GPP urban macro (UMa) non-line-of-sight (NLOS) scenarios, and adapt the QuaDRiGa simulator in accordance with the procedures proposed in the standardization discussions of 3GPP to enable the cluster-level SnS modeling, which is consistent with industrial measurements \cite{3gppr12410492}.
Unless specified otherwise, the simulation configurations follow the system model in \secref{sec:system_model}, with parameters summarized in \tabref{tab:simulation_configuration}. 

With these system configurations and the normalized maximum amplitude variation of ${\zeta} = 0.99$, the approximations—neglecting amplitude variation and high-order phase terms in NF propagation—is valid when the propagation distance exceeds $4.46$ m and $2.53$ m, respectively.

\subsubsection{Benchmarks and Performance Metric}
To demonstrate the superiority of the proposed algorithm, we select the following algorithms as benchmarks:
\begin{itemize}
  \item \textbf{VKF} \cite{kim2020massive}: Estimates spatial domain channels by the least square (LS) algorithm, with temporal correlations captured by the AR-based vector Kalman filter.
  \item \textbf{FIT} \cite{peng2019downlink}: Estimates beam-delay domain channels by the alternating LS (ALS) algorithm, with temporal correlations captured by the first-order Taylor series.
  \item \textbf{PAD} \cite{yin2020addressing}, \textbf{WTMP} \cite{li2024wavefront}: Extract dominant beam-delay domain taps by orthogonal matching pursuit (OMP) algorithm, with temporal correlations captured by Prony and matrix pencil methods for these taps.
\end{itemize}
Among these benchmarks, \textbf{VKF} and \textbf{FIT} are independent of specific beam domain structures, allowing them to be compatible with XL-MIMO systems.
Besides, \textbf{PAD} serves as the representative state-of-the-art channel prediction algorithms for massive MIMO systems, while \textbf{WTMP} further incorporates the NF propagation into the beam domain through the WT matrices.
For benchmarks that only predict channels for the future pilot symbols, we predict channels on the non-pilot symbols through MMSE interpolation, with prior knowledge of the maximum Doppler frequency and transmission power.
The normalized mean square error (NMSE) of SFT domain channels is adopted as the performance metric, defined by
\begin{equation}
    \text{NMSE} = \frac{\|\hat{\boldsymbol{\mathcal{H}}}^{\text{CP}} - \boldsymbol{\mathcal{H}}^{\text{CP}}\|_{F}^{2}}{\|\boldsymbol{\mathcal{H}}^{\text{CP}}\|_{F}^{2}},
\end{equation}
where $\boldsymbol{\mathcal{H}}^{\text{CP}}$ and $\hat{\boldsymbol{\mathcal{H}}}^{\text{CP}}$ denote the groud-truth and predicted SFT domain channels of upcoming OFDM symbols, respectively. 

\begin{table}[!t]
  \caption{Computational Complexity}
  \label{tab:computational_complexity}
  \centering
  \begin{tabular}{cc}
  \toprule
  \textbf{Algorithm} & \textbf{Computational Complexity} \\
  \midrule
  TS-BCP (per iteration) & $\mathcal{O}(N_\text{an}N_\text{sc}N_\text{sym}\bar{N})$ \\
  VKF &  $\mathcal{O}(N_\text{an}^{3}P^{3})$ \\
  FIT &  $\mathcal{O}(TRN_\text{an}N_\text{sc}N_\text{sym})$ \\
  PAD, WTMP &  $\mathcal{O}(TN_\text{an}N_\text{sc}N_\text{sym}(N_\text{an}F_\text{NF} + N_\text{sc} + T))$ \\
  \bottomrule
  \end{tabular}
\end{table}

\subsection{Computational Complexity}\label{sec:complexity}
The dominant computational complexity of \alref{alg:TSBCP} arises from multi-mode tensor-matrix multiplications, where computational complexity of each mode is equivalent to that of the corresponding matricized counterparts.
Notably, the computational complexity of multi-mode tensor-matrix multiplications depends on the execution order of modes: an order that minimizes the size of intermediate tensors yields lower computational complexity. 
Therefore, the tensor-matrix multiplications required for the SFT-to-BDD domain transformation are executed sequentially along the delay (frequency), beam (spatial), and Doppler (temporal) modes, while the inverse transformation adopts the reverse order.
To highlight the computational gains of tensor-matrix multiplications over matrix-vector multiplications, we assume that each dimension of BDD and SFT domain channels is of the same order, as is typically the case in practice.
As such, the computational complexity simplifies to $\mathcal{O}(N_\text{an}N_\text{sc}N_\text{sym}\bar{N})$, where $\bar{N} {\;\triangleq\;} N_\text{an} + N_\text{sc} + N_\text{sym}$ denotes the aggregate dimension of SFT domain channels. 
By comparison, the per-iteration computational complexity will increase to $\mathcal{O}(N_\text{an}^2 N_\text{sc}^2 N_\text{sym}^2)$ for matrix-vector multiplication without exploiting the separable structure of factor matrices.

\begin{figure}[!t]
    \centering
    \includegraphics[width = \scaleofsimulation \linewidth]{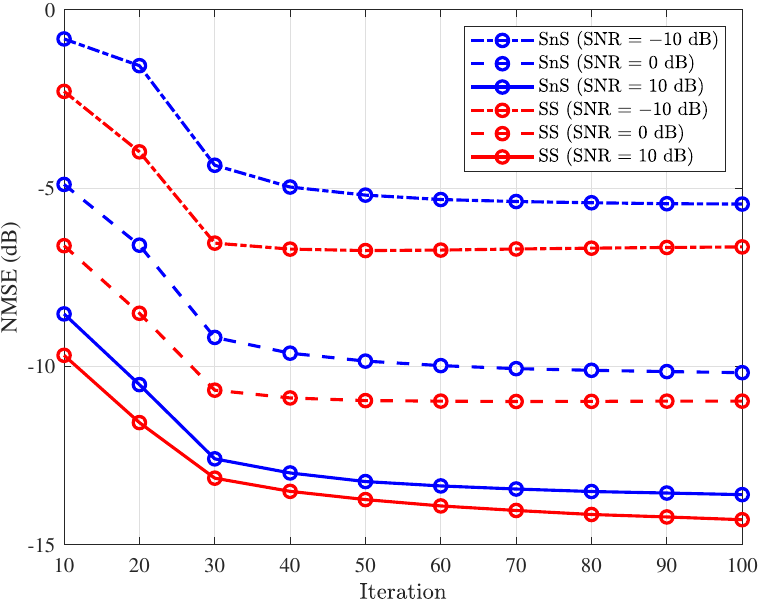}
    \caption{The convergence performance of the proposed algorithm.}
    \label{fig:Iteration}
\end{figure}

In the benchmarks, the computational complexity of both \textbf{PAD} and \textbf{WTMP} is dominated by the OMP algorithm, which can also exploit the separable structure of spatial-frequency domain factor matrices. Specifically, the computational complexity scales as $\mathcal{O}(TN_\text{an}N_\text{sc}N_\text{sym}(N_\text{an}F_\text{NF} + N_\text{sc} + T))$, with $F_\text{NF} = 1$ for \textbf{PAD} and $F_\text{NF} > 1$ for \textbf{WTMP} due to increased beam domain sampling induced by NF propagation.
For \textbf{VKF}, the critical step is the learning of AR parameters shared across all pilot subcarriers, yielding the computational complexity of $\mathcal{O}(P^{3}N_\text{an}^{3})$, where $P$ denotes the AR order, typically on the same order as the pilot OFDM symbols. 
In the case of \textbf{FIT}, its computational complexity is controlled by the ALS algorithm with $\mathcal{O}(TRN_\text{an}N_\text{sc}N_\text{sym})$, where $T$ denotes the number of iterations and $R$ denotes the predefined tensor rank, typically set as the maximum number of possible paths in the given environment. The computational complexity of the proposed algorithm and benchmarks are summarized in \tabref{tab:computational_complexity}.

\subsection{Performance Evaluation}
\subsubsection{Convergence Behavior}
To evaluate the convergence behavior of the proposed algorithm, we present the NMSE versus the number of iterations under different signal-to-noise ratio (SNR) levels in \figref{fig:Iteration}.
It is noteworthy that approximately $30$ iterations are sufficient to achieve nearly the full performance gains across various SNRs, while additional iterations result in only marginal refinements.
Due to the rapid decrease in NMSE during the initial iterations, early termination can be adopted to further reduce computational complexity, depending on the performance requirements of practical systems.
Across the full SNR range, the channel prediction performance without SnS propagation slightly outperforms that with SnS propagation.

\begin{figure}[!t]
  \centering
  \subfloat[]{
    \includegraphics[width = \scaleofsimulation \linewidth]{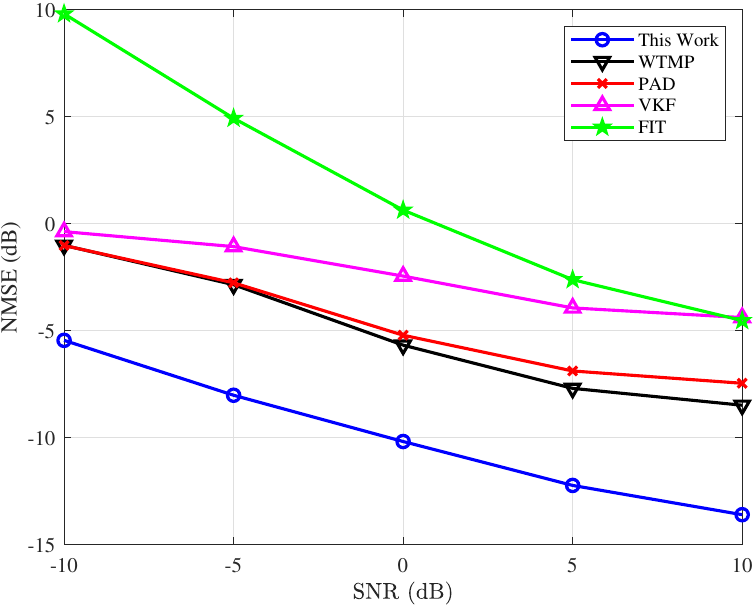}
    \label{fig:SNR_SnS}
  }
  \\
  \subfloat[]{
    \includegraphics[width = \scaleofsimulation \linewidth]{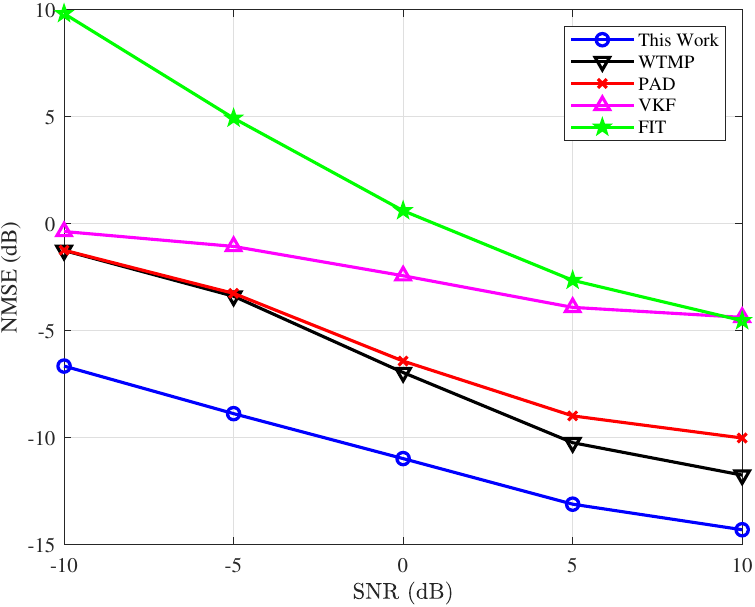}%
    \label{fig:SNR_SS}
  }
  \caption{NMSE of channel prediction versus SNR: (a) with SnS propagation effect, (b) without SnS propagation effect.}
  \label{fig:SNR}
\end{figure}

\subsubsection{NMSE versus SNR}
The NMSE of the proposed algorithm and benchmarks versus SNR is shown in \figref{fig:SNR}. 
Since \textbf{FIT} and \textbf{VKF} are independent of beam domain structure, they exhibit nearly identical channel prediction performance with and without SnS propagation.
However, the absence of beam domain modeling results in the poorest channel prediction performance, while the \textbf{FIT} algorithm further suffers from substantial degradation in low-SNR regime.
Owing to the incorporation of beam  and Doppler domain modeling, both \textbf{PAD} and \textbf{WTMP} achieve superior performance among the benchmarks, where \textbf{WTMP} benefits from spherical wave modeling and thus outperforms \textbf{PAD}.
Under SnS propagation, both \textbf{PAD} and \textbf{WTMP} exhibit the performance degradation of over $2$ dB compared to the case without SnS propagation at an SNR of $10$ dB, indicating that SnS propagation disrupts the beam domain representation.
The proposed algorithm significantly outperforms all the benchmarks, highlighting the performance gains brought by Doppler domain and SnS propagation modeling in channel prediction.
For instance, at an SNR of $10$ dB, it achieves at least $5$ dB gain in the presence of SnS propagation and $2$ dB in its absence, with even more substantial benefits observed in the low-SNR regime.

\begin{figure}[!t]
  \centering
  \subfloat[]{
    \includegraphics[width = \scaleofsimulation \linewidth]{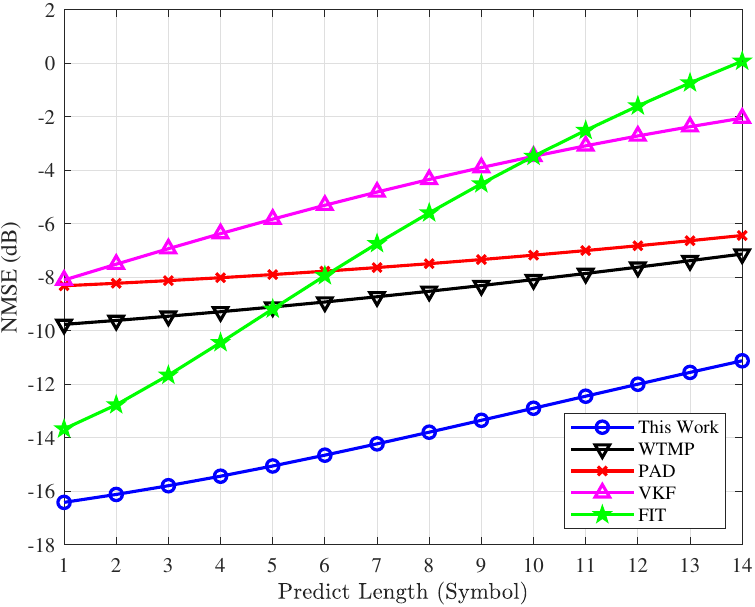}
    \label{fig:PredictLength_SnS}
  }
  \\
  \subfloat[]{
    \includegraphics[width = \scaleofsimulation \linewidth]{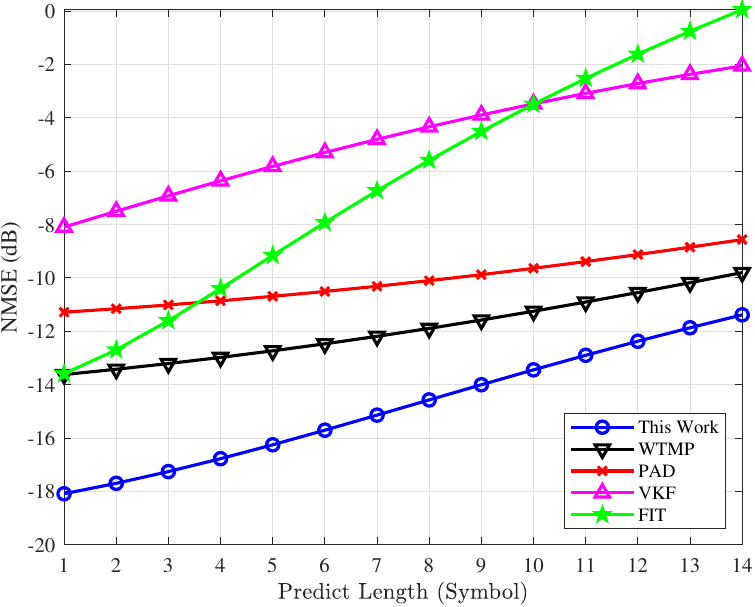}%
    \label{fig:PredictLength_SS}
  }
  \caption{NMSE of channel prediction versus prediction length at $\text{SNR} = 10$ dB: (a) with SnS propagation effect, (b) without SnS propagation effect.}
  \label{fig:PredictLength}
\end{figure}

\subsubsection{NMSE versus Prediction Length}
To illustrate the variation of channel prediction performance with prediction length, we present the NMSEs for symbols between the current pilot symbols and the first future pilot symbols in \figref{fig:PredictLength}. 
As expected, the performance of both the proposed algorithm and the benchmarks degrades as the prediction length increases, consistent with prior studies on channel prediction.
Although \textbf{FIT} outperforms all benchmarks at shorter prediction lengths under SnS propagation, its performance degrades rapidly by approximately $14$ dB, due to the limited accuracy of the first-order Taylor series in modeling temporal correlations.
In contrast, \textbf{VKF}, which employs AR-based temporal correlation modeling, achieves better long-term predictions than \textbf{FIT}, though it still exhibits about $6$ dB degradation as prediction length increases.
Both \textbf{PAD} and \textbf{WTMP} demonstrate improved stability across prediction lengths, indicating that the Prony and matrix pencil methods offer more accurate temporal correlation modeling than both the first-order Taylor series and AR models.
In comparison, the proposed algorithm consistently outperforms all benchmark methods across all prediction lengths.
For the first future non-pilot and pilot symbols, it achieves NMSEs below $-16$ dB and $-11$ dB, respectively, with SnS propagation effects, and approximately $-18$ dB and $-11.5$ dB without them.

\subsubsection{NMSE versus Carrier Frequency}
\begin{figure}[!t]
  \centering
  \subfloat[]{
    \includegraphics[width = \scaleofsimulation \linewidth]{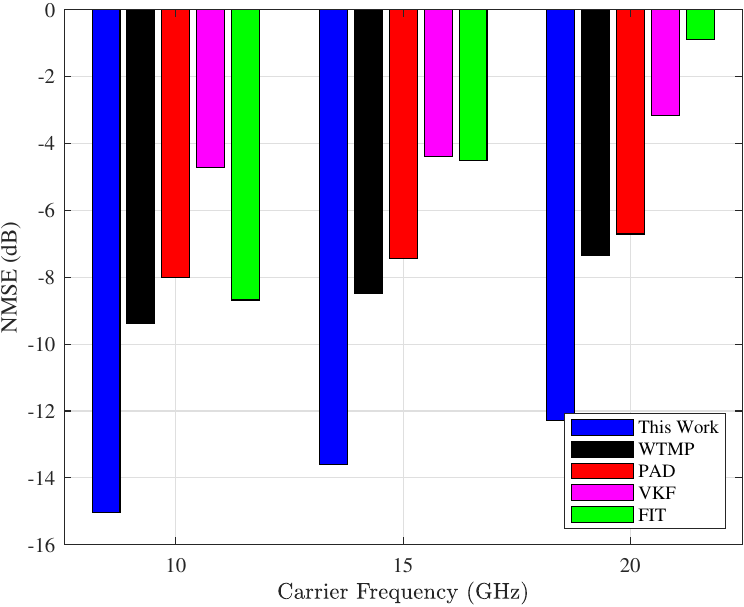}
    \label{fig:CarrierFreq_SnS}
  }
  \\
  \subfloat[]{
    \includegraphics[width = \scaleofsimulation \linewidth]{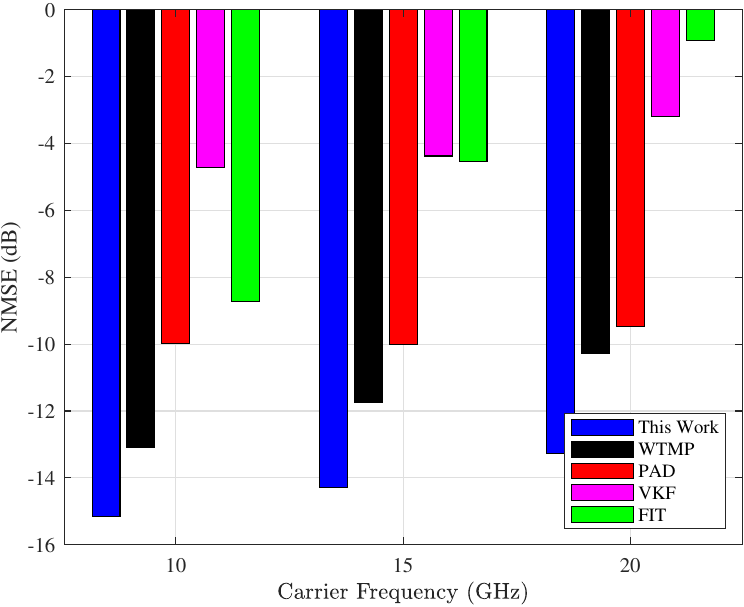}%
    \label{fig:CarrierFreq_SS}
  }
  \caption{NMSE of channel prediction versus carrier frequencies at $\text{SNR} = 10$ dB: (a) with SnS propagation effect, (b) without SnS propagation effect.}
  \label{fig:CarrierFreq}
\end{figure}

Along with the typical carrier frequency of $15$ GHz, we also evaluate the channel prediction performance of the proposed algorithm at both higher ($20$ GHz) and lower ($10$ GHz) frequencies within the upper mid-band, as shown in \figref{fig:CarrierFreq}.
Due to the linear dependence of Doppler frequency on carrier frequency, all benchmarks as well as the proposed algorithm suffer from performance degradation as the carrier frequency grows, while the severe performance drop of approximately $9$ dB observed for \textbf{FIT} further highlights its limited capability in capturing temporal correlations.
In contrast, the other benchmarks and the proposed algorithm maintain relatively stable performance over these carrier frequencies.
An interesting observation is that the performance gap between \textbf{PAD} and \textbf{WTMP} without SnS propagation shrinks from over $3$ dB to approximately $0.5$ dB as the carrier frequency increases. 
This is due to the reduction in Rayleigh distance at higher carrier frequencies when the number of antennas remains fixed, which weakens NF propagation and consequently reduces the performance difference between \textbf{PAD} and \textbf{WTMP}.
Across all carrier frequencies, the proposed algorithm consistently delivers the best and most robust channel prediction performance, demonstrating its effectiveness for upper mid-band systems.

\section{Conclusion}\label{sec:conclusion}
This paper investigated tensor-structured Bayesian channel prediction for upper mid-band XL-MIMO systems, explicitly addressing the challenges posed by NF and SnS propagation.
By developing multi-linear SFT domain channel models and probabilistic models with perturbation-aware BDD sampling, the proposed method effectively captures the inherent sparsity of channels and facilitates physics-inspired channel prediction.
Following the probabilistic models, we formulated the channel prediction under the MMSE criterion with unknown model hyperparameters, and developed the TS-BLI algorithm within the EM framework for channel prediction.
In this algorithm, the multi-linear structure of channels allows for the bi-layer factor graph representation and facilitates tensor operations that reduce computational complexity.
Simulation results demonstrate that the proposed algorithm significantly outperforms existing benchmarks in channel prediction accuracy, underscoring its potential for practical deployment in future large-scale wireless communication systems.


\ifCLASSOPTIONcaptionsoff
  \newpage
\fi



\bibliographystyle{IEEEtran}
\bibliography{IEEEabrv, reference}

\begin{thebibliography}{10}
\providecommand{\url}[1]{#1}
\csname url@samestyle\endcsname
\providecommand{\newblock}{\relax}
\providecommand{\bibinfo}[2]{#2}
\providecommand{\BIBentrySTDinterwordspacing}{\spaceskip=0pt\relax}
\providecommand{\BIBentryALTinterwordstretchfactor}{4}
\providecommand{\BIBentryALTinterwordspacing}{\spaceskip=\fontdimen2\font plus
\BIBentryALTinterwordstretchfactor\fontdimen3\font minus \fontdimen4\font\relax}
\providecommand{\BIBforeignlanguage}[2]{{%
\expandafter\ifx\csname l@#1\endcsname\relax
\typeout{** WARNING: IEEEtran.bst: No hyphenation pattern has been}%
\typeout{** loaded for the language `#1'. Using the pattern for}%
\typeout{** the default language instead.}%
\else
\language=\csname l@#1\endcsname
\fi
#2}}
\providecommand{\BIBdecl}{\relax}
\BIBdecl

\bibitem{giordani2020toward}
M.~Giordani, M.~Polese, M.~Mezzavilla, S.~Rangan, and M.~Zorzi, ``Toward {6G} networks: Use cases and technologies,'' \emph{{IEEE} Commun. Mag.}, vol.~58, no.~3, pp. 55--61, Mar. 2020.

\bibitem{na2024operator}
M.~Na, J.~Lee, G.~Choi, T.~Yu, J.~Choi, J.~Lee, and S.~Bahk, ``Operator's perspective on {6G}: {6G} services, vision, and spectrum,'' \emph{{IEEE} Commun. Mag.}, vol.~62, no.~8, pp. 178--184, Aug. 2024.

\bibitem{wang2024towards}
Y.~Wang, H.~Hou, X.~Yi, W.~Wang, and S.~Jin, ``Towards unified {AI} models for {MU-MIMO} communications: A tensor equivariance framework,'' \emph{arXiv preprint arXiv:2406.09022}, 2024.

\bibitem{3gpp38820}
\BIBentryALTinterwordspacing
{3rd Generation Partnership Project (3GPP)}, ``Study on the 7 to 24 {GHz} frequency range for {NR},'' {3GPP}, Technical Specification (TS) 38.820, 2020. [Online]. Available: \url{https://www.3gpp.org}
\BIBentrySTDinterwordspacing

\bibitem{kang2024cellular}
S.~Kang, M.~Mezzavilla, S.~Rangan, A.~Madanayake, S.~B. Venkatakrishnan, G.~Hellbourg, M.~Ghosh, H.~Rahmani, and A.~Dhananjay, ``Cellular wireless networks in the upper mid-band,'' \emph{IEEE Open J. Commun. Soc.}, vol.~5, pp. 2058--2075, Mar. 2024.

\bibitem{tian2025mid}
J.~Tian, Y.~Han, X.~Li, S.~Jin, and C.-K. Wen, ``Mid-band extra large-scale {MIMO} system: Channel modeling and performance analysis,'' \emph{{IEEE} Trans. Commun.}, vol.~73, no.~2, pp. 1025--1041, Feb. 2025.

\bibitem{miao2023sub}
H.~Miao, J.~Zhang, P.~Tang, L.~Tian, X.~Zhao, B.~Guo, and G.~Liu, ``Sub-6 {GHz} to {mmWave} for {5G}-advanced and beyond: Channel measurements, characteristics and impact on system performance,'' \emph{{IEEE} J. Sel. Areas Commun.}, vol.~41, no.~6, pp. 1945--1960, Jun. 2023.

\bibitem{yuan2023spatial}
Z.~Yuan, J.~Zhang, Y.~Ji, G.~F. Pedersen, and W.~Fan, ``Spatial non-stationary near-field channel modeling and validation for massive {MIMO} systems,'' \emph{{IEEE} Trans. Wireless Commun.}, vol.~71, no.~1, pp. 921--933, Jan. 2023.

\bibitem{guillaud2004specular}
M.~Guillaud and D.~T. Slock, ``A specular approach to {MIMO} frequency-selective channel tracking and prediction,'' in \emph{Proc. IEEE Signal Process. Advances in Wireless Commun.}, 2004, pp. 59--63.

\bibitem{adeogun2015extrapolation}
R.~O. Adeogun, P.~D. Teal, and P.~A. Dmochowski, ``Extrapolation of {MIMO} mobile-to-mobile wireless channels using parametric-model-based prediction,'' \emph{{IEEE} Trans. Veh. Technol.}, vol.~64, no.~10, pp. 4487--4498, Oct. 2015.

\bibitem{yin2020addressing}
H.~Yin, H.~Wang, Y.~Liu, and D.~Gesbert, ``Addressing the curse of mobility in massive {MIMO} with prony-based angular-delay domain channel predictions,'' \emph{{IEEE} J. Sel. Areas Commun.}, vol.~38, no.~12, pp. 2903--2917, Dec. 2020.

\bibitem{qin2022partial}
Z.~Qin, H.~Yin, Y.~Cao, W.~Li, and D.~Gesbert, ``A partial reciprocity-based channel prediction framework for {FDD} massive {MIMO} with high mobility,'' \emph{{IEEE} Trans. Wireless Commun.}, vol.~21, no.~11, pp. 9638--9652, Nov. 2022.

\bibitem{li2022multi}
W.~Li, H.~Yin, Z.~Qin, Y.~Cao, and M.~Debbah, ``A multi-dimensional matrix pencil-based channel prediction method for massive {MIMO} with mobility,'' \emph{{IEEE} Trans. Wireless Commun.}, vol.~22, no.~4, pp. 2215--2230, Apr. 2023.

\bibitem{wu2021channel}
C.~Wu, X.~Yi, Y.~Zhu, W.~Wang, L.~You, and X.~Gao, ``Channel prediction in high-mobility massive {MIMO}: From spatio-temporal autoregression to deep learning,'' \emph{{IEEE} J. Sel. Areas Commun.}, vol.~39, no.~7, pp. 1915--1930, Jul. 2021.

\bibitem{wang2023channel}
X.~Wang, Y.~Shi, W.~Xin, T.~Wang, G.~Yang, and Z.~Jiang, ``Channel prediction with time-varying {Doppler} spectrum in high-mobility scenarios: A polynomial {Fourier} transform based approach and field measurements,'' \emph{{IEEE} Trans. Wireless Commun.}, vol.~22, no.~11, pp. 7116--7129, Nov. 2023.

\bibitem{wan2023robust}
Y.~Wan, G.~Liu, A.~Liu, and M.-J. Zhao, ``Robust multi-user channel tracking scheme for {5G} new radio,'' \emph{{IEEE} Trans. Wireless Commun.}, 2023.

\bibitem{wan2024two}
Y.~Wan and A.~Liu, ``A two-stage {2D} channel extrapolation scheme for {TDD} {5G} {NR} systems,'' \emph{{IEEE} Trans. Wireless Commun.}, 2024.

\bibitem{zhu2024joint}
Y.~Zhu, J.~Zhuang, G.~Sun, H.~Hou, L.~You, and W.~Wang, ``Joint channel estimation and prediction for massive {MIMO} with frequency hopping sounding,'' \emph{{IEEE} Trans. Wireless Commun.}, 2024.

\bibitem{shi2022channel}
D.~Shi, L.~Song, W.~Zhou, X.~Gao, C.-X. Wang, and G.~Y. Li, ``Channel acquisition for {HF} skywave massive {MIMO}-{OFDM} communications,'' \emph{{IEEE} Trans. Wireless Commun.}, vol.~22, no.~6, pp. 4074--4089, Jun. 2023.

\bibitem{hou2024tensor}
H.~Hou, Y.~Wang, Y.~Zhu, X.~Yi, W.~Wang, D.~Slock, and S.~Jin, ``A tensor-structured approach to dynamic channel prediction for massive {MIMO} systems with temporal non-stationarity,'' \emph{arXiv preprint arXiv:2412.06713}, 2024.

\bibitem{cui2022channel}
M.~Cui and L.~Dai, ``Channel estimation for extremely large-scale {MIMO}: Far-field or near-field?'' \emph{{IEEE} Trans. Commun.}, vol.~70, no.~4, pp. 2663--2677, Jan. 2022.

\bibitem{lu2023near}
Y.~Lu and L.~Dai, ``Near-field channel estimation in mixed {LoS}/{NLoS} environments for extremely large-scale {MIMO} systems,'' \emph{{IEEE} Trans. Commun.}, vol.~71, no.~6, pp. 3694--3707, Jun. 2023.

\bibitem{9940281}
Z.~Hu, C.~Chen, Y.~Jin, L.~Zhou, and Q.~Wei, ``Hybrid-field channel estimation for extremely large-scale massive {MIMO} system,'' \emph{{IEEE} Commun. Lett.}, vol.~27, no.~1, pp. 303--307, Jan. 2023.

\bibitem{9598863}
X.~Wei and L.~Dai, ``Channel estimation for extremely large-scale massive {MIMO}: Far-field, near-field, or hybrid-field?'' \emph{{IEEE} Commun. Lett.}, vol.~26, no.~1, pp. 177--181, Jan. 2022.

\bibitem{hou2024beam}
H.~Hou, X.~He, T.~Fang, X.~Yi, W.~Wang, and S.~Jin, ``Beam-delay domain channel estimation for {mmWave} {XL}-{MIMO} systems,'' \emph{{IEEE} J. Sel. Topics Signal Process.}, vol.~18, no.~4, pp. 646--661, May 2024.

\bibitem{han2020channel}
Y.~Han, S.~Jin, C.-K. Wen, and X.~Ma, ``Channel estimation for extremely large-scale massive {MIMO} systems,'' \emph{{IEEE} Wireless Commun. Lett.}, vol.~9, no.~5, pp. 633--637, May 2020.

\bibitem{chen2024non}
Y.~Chen and L.~Dai, ``Non-stationary channel estimation for extremely large-scale {MIMO},'' \emph{{IEEE} Trans. Wireless Commun.}, vol.~23, no.~7, pp. 7683--7697, Jul. 2024.

\bibitem{yang2025adaptive}
S.~Yang, P.~An, P.~Yang, X.~Cao, D.~O. Wu, and T.~Q. Quek, ``Adaptive subarray segmentation: A new paradigm of spatial non-stationary near-field channel estimation for {XL}-{MIMO} systems,'' \emph{arXiv preprint arXiv:2503.04211}, 2025.

\bibitem{zhu2021bayesian}
Y.~Zhu, H.~Guo, and V.~K. Lau, ``Bayesian channel estimation in multi-user massive {MIMO} with extremely large antenna array,'' \emph{{IEEE} Trans. Signal Process.}, vol.~69, pp. 5463--5478, Sep. 2021.

\bibitem{tang2024joint}
A.~Tang, J.-B. Wang, Y.~Pan, W.~Zhang, X.~Zhang, Y.~Chen, H.~Yu, and R.~C. De~Lamare, ``Joint visibility region and channel estimation for extremely large-scale {MIMO} systems,'' \emph{{IEEE} Trans. Commun.}, vol.~72, no.~10, pp. 6087--6101, Oct. 2024.

\bibitem{xu2024joint}
W.~Xu, A.~Liu, M.-j. Zhao, and G.~Caire, ``Joint visibility region detection and channel estimation for {XL}-{MIMO} systems via alternating {MAP},'' \emph{{IEEE} Trans. Signal Process.}, vol.~72, pp. 4827--4842, Oct. 2024.

\bibitem{xu2025exploiting}
W.~Xu, A.~Liu, M.~jian Zhao, G.~Caire, and Y.-C. Wu, ``Exploiting dynamic sparsity for near-field spatial non-stationary {XL}-{MIMO} channel tracking,'' \emph{arXiv preprint arXiv:2412.19475}, 2025.

\bibitem{tang2024spatially}
A.~Tang, J.-B. Wang, Y.~Pan, W.~Zhang, Y.~Chen, H.~Yu, and R.~C. de~Lamare, ``Spatially non-stationary {XL}-{MIMO} channel estimation: A three-layer generalized approximate message passing method,'' \emph{{IEEE} Trans. Signal Process.}, vol.~73, pp. 356--371, Dec. 2024.

\bibitem{li2024wavefront}
W.~Li, H.~Yin, Z.~Qin, and M.~Debbah, ``Wavefront transformation-based near-field channel prediction for extremely large antenna array with mobility,'' \emph{{IEEE} Trans. Wireless Commun.}, vol.~23, no.~10, pp. 15\,613--15\,626, Oct. 2024.

\bibitem{kolda2009tensor}
T.~G. Kolda and B.~W. Bader, ``Tensor decompositions and applications,'' \emph{SIAM Review}, vol.~51, no.~3, pp. 455--500, 2009.

\bibitem{brazell2013solving}
M.~Brazell, N.~Li, C.~Navasca, and C.~Tamon, ``Solving multilinear systems via tensor inversion,'' \emph{SIAM J. Matrix Anal. Appl.}, vol.~34, no.~2, pp. 542--570, 2013.

\bibitem{3gpp38211}
\BIBentryALTinterwordspacing
{3rd Generation Partnership Project (3GPP)}, ``{NR}; physical channels and modulation,'' {3GPP}, Technical Specification (TS) 38.211, 2025. [Online]. Available: \url{https://www.3gpp.org}
\BIBentrySTDinterwordspacing

\bibitem{sherman1962properties}
J.~Sherman, ``Properties of focused apertures in the {Fresnel} region,'' \emph{IRE {Trans}. {Antennas} {Propag}.}, vol.~10, no.~4, pp. 399--408, Jul. 1962.

\bibitem{3gppr12406199}
R1-2406199, ``Views on channel model adaptation/extension of {TR}38.901 for 7-24{GHz},'' {Vivo}, 3GPP TSG RAN WG1 Meeting 118, 2024.

\bibitem{selvan2017fraunhofer}
K.~T. Selvan and R.~Janaswamy, ``Fraunhofer and {Fresnel} distances: Unified derivation for aperture antennas,'' \emph{{IEEE} Antennas Propag. Mag.}, vol.~59, no.~4, pp. 12--15, Aug. 2017.

\bibitem{cheng2022towards}
L.~Cheng, Z.~Chen, Q.~Shi, Y.-C. Wu, and S.~Theodoridis, ``Towards flexible sparsity-aware modeling: Automatic tensor rank learning using the generalized hyperbolic prior,'' \emph{{IEEE} Trans. Signal Process.}, vol.~70, pp. 1834--1849, Apr. 2022.

\bibitem{moon1996expectation}
T.~K. Moon, ``The expectation-maximization algorithm,'' \emph{{IEEE} Signal Process. Mag.}, vol.~13, no.~6, pp. 47--60, Nov. 1996.

\bibitem{zou2021multi}
Q.~Zou, H.~Zhang, and H.~Yang, ``Multi-layer bilinear generalized approximate message passing,'' \emph{{IEEE} Trans. Signal Process.}, vol.~69, pp. 4529--4543, Jul. 2021.

\bibitem{rangan2011generalized}
S.~Rangan, ``Generalized approximate message passing for estimation with random linear mixing,'' in \emph{Proc. IEEE Int. Symp. Inf. Theory}.\hskip 1em plus 0.5em minus 0.4em\relax IEEE, Jul. 2011, pp. 2168--2172.

\bibitem{parker2014bilinear}
J.~T. Parker, P.~Schniter, and V.~Cevher, ``Bilinear generalized approximate message passing—{Part} {I}: Derivation,'' \emph{{IEEE} Trans. Signal Process.}, vol.~62, no.~22, pp. 5839--5853, Nov. 2014.

\bibitem{vila2013expectation}
J.~P. Vila and P.~Schniter, ``Expectation-maximization {Gaussian}-mixture approximate message passing,'' \emph{{IEEE} Trans. Signal Process.}, vol.~61, no.~19, pp. 4658--4672, Oct. 2013.

\bibitem{zhang2021unifying}
D.~Zhang, X.~Song, W.~Wang, G.~Fettweis, and X.~Gao, ``Unifying message passing algorithms under the framework of constrained {Bethe} free energy minimization,'' \emph{{IEEE} Trans. Wireless Commun.}, vol.~20, no.~7, pp. 4144--4158, Jul. 2021.

\bibitem{3gpp38901}
\BIBentryALTinterwordspacing
{3rd Generation Partnership Project (3GPP)}, ``Study on channel model for frequencies from 0.5 to 100 {GHz} ({Release 18}),'' {3GPP}, Technical Report (TR) 38.901, 2024. [Online]. Available: \url{https://www.3gpp.org}
\BIBentrySTDinterwordspacing

\bibitem{jaeckel2014quadriga}
S.~Jaeckel, L.~Raschkowski, K.~B{\"o}rner, and L.~Thiele, ``{QuaDRiGa}: A 3-{D} multi-cell channel model with time evolution for enabling virtual field trials,'' \emph{{IEEE} Trans. Antennas Propag.}, vol.~62, no.~6, pp. 3242--3256, Jun. 2014.

\bibitem{3gppr12410492}
R1-2410492, ``Channel model adaptation or extension of {TR}38.901 for 7-24{GHz},'' {Qualcomm}, 3GPP TSG RAN WG1 Meeting 119, 2024.

\bibitem{kim2020massive}
H.~Kim, S.~Kim, H.~Lee, C.~Jang, Y.~Choi, and J.~Choi, ``Massive {MIMO} channel prediction: Kalman filtering vs. machine learning,'' \emph{{IEEE} Trans. Commun.}, vol.~69, no.~1, pp. 518--528, Jan. 2021.

\bibitem{peng2019downlink}
W.~Peng, W.~Li, W.~Wang, X.~Wei, and T.~Jiang, ``Downlink channel prediction for time-varying {FDD} massive {MIMO} systems,'' \emph{{IEEE} J. Sel. Topics Signal Process.}, vol.~13, no.~5, pp. 1090--1102, Sep. 2019.

\end{thebibliography}
%

%








\end{document}